\documentclass[runningheads]{llncs}
\usepackage[margin=1in]{geometry}
\usepackage[T1]{fontenc}
\usepackage{hyperref}
\usepackage{amssymb}
\usepackage{amsmath}
\usepackage{float}
\usepackage{graphicx}
\usepackage{amsfonts}
\usepackage{amssymb}
\usepackage{paralist}
\usepackage{floatflt}
\usepackage{alltt}
\usepackage{color}

\usepackage[ruled,linesnumbered]{algorithm2e}
\usepackage{algorithmic}
\usepackage{subcaption}
\usepackage{tikz}
\usepackage{bm}
\usepackage{tabularx,environ}
\usetikzlibrary{calc}
\tikzset{
    tree/.style={thin, solid, draw=black},
    link/.style={thin, dashed, draw=black},
    path/.style={blue, loosely dotted},
    divide/.style={blue, densely dotted},
    standard/.style={thin, solid, draw=black},
    marked/.style={ultra thick, dashed, draw=red},
    semimarked/.style={very thick, dotted, draw=green},
    steiner/.style={thin, circle, solid, draw=black},
    terminal/.style={thin, rectangle, solid, draw=black},
    goodfather/.style={postaction={decorate},
        decoration={markings,mark=at position .55 with {\arrow[scale=3, draw=blue]{>}}}},
    leaf/.style={circle, thin, fill=none, draw=black},
    subroot/.style={minimum size = 1cm, rectangle, thin, fill=none, draw=black},
}

\newtheorem{claims}{Claim}

\begin{document}

\title{A Global Analysis of the Primal-Dual Method for Edge Augmentation Problems}
\author{Ishan Bansal}
\institute{Amazon, Bellevue, WA, USA}

\maketitle
\begin{abstract}
    We study a core algorithmic problem in network design called $\mathcal{F}$-augmentation that involves increasing the connectivity of a given family of cuts $\mathcal{F}$. Over 30 years ago, Williamson et al. (STOC `93) provided a 2-approximation primal-dual algorithm when $\mathcal{F}$ is a so-called uncrossable family but extending their results to families that are non-uncrossable has remained a challenging question.

    In this paper, we introduce the novel concept of the crossing density of a set family and show how this opens up a completely new approach to analyzing primal-dual algorithms. We study pliable families, a strict generalization of uncrossable families introduced by Bansal et al. (ICALP `23), and provide the first approximation algorithm for $\mathcal{F}$-augmentation of general pliable families.
    
    We also improve on the results in Bansal et al. (ICALP `23) by providing a 6-approximation algorithm for the $\mathcal{F}$-augmentation problem when $\mathcal{F}$ is a family of near min-cuts. This immediately improves approximation factors for the Capacitated Network Design Problem.

    Finally, we study the $(p,3)$-flexible graph connectivity problem. By carefully analyzing the structure of feasible solutions and using the techniques developed in this paper, we provide the first constant factor approximation algorithm for this problem exhibiting an 12-approximation algorithm.
\end{abstract}

\begin{remark}
    We are reverting back to an earlier version of this paper (v2) published in June 2024 that already contained the 6-approximation algorithm and its proof (see Theorem 3 in (v2)). Earlier versions of this paper also claimed a 5-approximation but there was a bug in the proof. In particular, Theorem 5 in (v3) is incorrect. Recently (in April 2025), Nutov \cite{nutov2025tightanalysisprimaldualmethod} posted a 6-approximation algorithm, with a very similar proof to the one published in (v2) of this paper. \cite{nutov2025tightanalysisprimaldualmethod} does contain other novel and interesting results.
\end{remark}

\section{Introduction}

A branch of network design involves finding the optimal cost of maintaining and designing networks to meet certain connectivity requirements. This area has been studied for decades and has found applications in transportation, supply-chain, security networks, fleet management etc. A well known and widely used algorithmic idea to design approximation algorithms for such network design problems is that of edge augmentation - instead of finding a network that immediately meets all the given connectivity requirements, we start with an infeasible solution and increase the connectivity of this solution one step at a time. As an example, consider the $k$-edge connected spanning subgraph problem. Here, one could start with the empty set and find a 1-edge connected subgraph first (this is just the minimum spanning tree problem). Then, one could augment this spanning tree to make it $2$-edge connected (this is known as the tree augmentation problem), and further one could solve the resulting weighted connectivity augmentation problems to obtain a $3,4,\ldots, k$-edge connected spanning subgraph. 

This methodology of gradually increasing connectivity has been used for decades starting from classical $k$-edge/node connectivity and survivable network design problems \cite{williamson1993primal,goemans1994improved,ravi1997approximation} to more recent non-uniform fault tolerant models like flexible graph connectivity and bulk-robust network design problems \cite{adjiashvili2022flexible,boyd2024approximation,bansal2024improved,nutov2023improved,nutov2024improved,chekuri2023}. Furthermore, for some of these problems like flexible graph connectivity, this augmentation methodology is currently the only known technique to obtain good approximation algorithms. A formal description of this augmentation framework is given by the $\bm{\mathcal{F}}$\textbf{-augmentation} problem\footnote{$\mathcal{F}$-augmentation has also been referred to as the 0-1 $f$-connectivity problem in the literature}. A good algorithmic understanding of the $\mathcal{F}$-augmentation problem is evidently extremely useful to the design of good approximation algorithms for a wide variety of network design problems.

\begin{problem}[$\mathcal{F}$-augmentation] ~\\ \textit{Input:} Graph $G=(V,E)$ with edge costs $\{c_e\}_{e\in E}$ and a family of cuts $\mathcal{F} \subseteq 2^V$. \\ \textit{Output:} Subset of the edges $F\subseteq E$ with minimum $c(F) = \sum_{e\in F}c_e$ such that for every $S\in \mathcal{F}$, the set of edges $F\cap\delta(S)$\footnote{$\delta(S)$ refers to the set of edges with exactly one en point in the set $S$} is non-empty.
\end{problem}

A widely used and versatile tool to tackle the $\mathcal{F}$-augmentation problem is the primal-dual method. These are methods that are based on linear programming and utilize the dual linear program to guide the algorithm in finding a feasible and close to optimal primal solution. In the context of edge-augmentation, the primal-dual method has been used to tackle various problems like generalized Steiner tree \cite{agrawal1991trees}, node connectivity \cite{ravi1997approximation}, element connectivity \cite{jain2002primal}, survivable network design \cite{goemans1994improved}, flexible graph connectivity \cite{bansal2024improved}, capacitated network design \cite{bansal2024improved}, directed vertex connectivity augmentation \cite{vegh2008primal}, directed network design \cite{melkonian2005primal}, and many more. In this paper we focus on the analysis of the primal-dual method on the $\mathcal{F}$-augmentation problem going beyond previous literature. As a byproduct of our work, we are able to provide improved approximation algorithms for certain concrete network design problems like capacitated-$k$-edge connectivity and flexible graph connectivity.

The $\mathcal{F}$-augmentation problem captures the hitting set problem \cite{bansal2024improved}. Unfortunately, this implies that unless $P=NP$, it is impossible to find an approximation algorithm for the problem with approximation factor better than $O(\log |V|)$. Thus, one needs to exploit underlying structures of the family of cuts $\mathcal{F}$ to be able to design good approximation algorithms. In a seminal paper, Williamson, Goemans, Mihail and Vazirani (WGMV) \cite{williamson1993primal} considered a versatile property for a family of cuts $\mathcal{F}$ that in particular captures the family of minimum cuts of any graph $G$. They called such families of cuts \textbf{uncrossable}.

\begin{definition}[Uncrossable Family]\label{def:uncrossable}
    A family of sets $\mathcal{F}$ is called uncrossable if $A,B \in \mathcal{F}$ implies that either (i) $A\cup B \in \mathcal{F}$ and $A\cap B \in \mathcal{F}$ or (ii) $A\setminus B \in \mathcal{F}$ and $B\setminus A \in \mathcal{F}$.
\end{definition}

Building on the works of \cite{adjiashvili2022flexible,goemans1995general}, they provided a primal-dual 2-approximation algorithm for the $\mathcal{F}$-augmentation problem when $\mathcal{F}$ is an uncrossable family. This tool developed by WGMV has been extremely valuable to the network design community, finding applications in the survivable network design problem \cite{goemans1994improved}, element connectivity problems \cite{jain2002primal}, flexible graph connectivity \cite{chekuri2023,boyd2024approximation,bansal2024improved,nutov2023improved}, capacitated network design \cite{bansal2024improved}. With certain tweaks, their primal-dual method has also been applied to node connectivity problems \cite{ravi1997approximation}, and directed network design problems \cite{vegh2008primal,melkonian2005primal,friggstad2023constant}. We will refer to their algorithm as the \textit{WGMV primal-dual algorithm} and a brief description of the algorithm is provided in Section \ref{sec:prelims}. In their paper, WGMV state ``Extending our algorithm to handle non-uncrossable
functions remains a challenging open problem''. Their challenge has stood the test of time since for nearly 30 years, no extensions of their algorithm beyond uncrossable functions were known. This gap in our understanding was not because interesting non-uncrossable families of cuts were unknown. In fact a very important and widely studied family of cuts do not satisfy the uncrossable property namely the family of $\alpha$-min cuts of a graph. A cut is an $\alpha$-min cut if its size is at most $\alpha$ times the size of a minimum cut. The structure of $\alpha$-min cuts has been extensively studied for decades and is a fundamental family of cuts in the area of network design \cite{Karger93,karger1996new,benczur2008deformable}.

Very recently, Bansal, Cheriyan, Grout and Ibrahimpur (BCGI) \cite{bansal2024improved} considered a natural generalization of uncrossable families that they called \textbf{pliable} families. For any two cuts $A,B\subseteq V$, the four cuts $A\cup B, A\cap B, A\setminus B, B\setminus A$ are referred to as the \textit{corner cuts}. While uncrossability requires that for $A,B \in \mathcal{F}$, one of two particular pairs of corner cuts also lie within $\mathcal{F}$, pliability relaxes this condition to allow for any pair of corner cuts to lie within $\mathcal{F}$.

\begin{definition}[Pliable Family]\label{def:pliable}
    A family of sets $\mathcal{F}$ is called pliable if $A,B \in \mathcal{F}$ implies that at least two of the four cuts $A\cup B, A\cap B, A\setminus B, B\setminus A$ also lie in $\mathcal{F}$.
\end{definition}

Pliable families generalize uncrossable families in the most natural way but also manage to capture the family of $\alpha$-min cuts. This is simple to see using the submodularity \eqref{eq:submodularity} and posimodularity \eqref{eq:posimodularity} of the cut function. Indeed if $A$ and $B$ are $\alpha$-min cuts, then $|\delta(A)| \leq \alpha\lambda$ and $|\delta(B)| \leq \alpha\lambda$ where $\lambda$ is the size of a minimum cut. Then, equation \eqref{eq:submodularity} tells us that at least one of $|\delta(A\cup B)|$ or $|\delta(A\cap B)|$ is at most $\alpha\lambda$ and similarly equation \eqref{eq:posimodularity} tells us that at least one of $|\delta(A\setminus B)|$ or $|\delta(B\setminus A)|$ is at most $\alpha\lambda$. Hence at least two of the four corner cuts are also $\alpha$-min cuts.
\begin{align}
    &|\delta(A)| + |\delta(B)| \geq |\delta(A\cup B)| + |\delta(A\cap B)| \label{eq:submodularity}\\
    &|\delta(A)| + |\delta(B)| \geq |\delta(A\setminus B)| + |\delta(B\setminus A)| \label{eq:posimodularity}
\end{align}

The main contribution of BCGI was to extend the WGMV primal-dual algorithm to work for a sub-family of pliable families that also captures the family of $\alpha$-min cuts. They called such families $\gamma$-pliable families and these were characterized by a local property of cuts (a technical definition is provided in the next section). They showed that the WGMV primal-dual algorithm is a 16-approximation algorithm for the $\mathcal{F}$-augmentation problem when $\mathcal{F}$ is a $\gamma$-pliable family. Subsequently, Nutov \cite{nutov2023extend} considered a further sub-family of $\gamma$-pliable families called semi-uncrossable families and showed that the WGMV primal-dual algorithm is a 2-approximation algorithm in this case.

\begin{figure}
    \centering
    \includegraphics[width=1\linewidth]{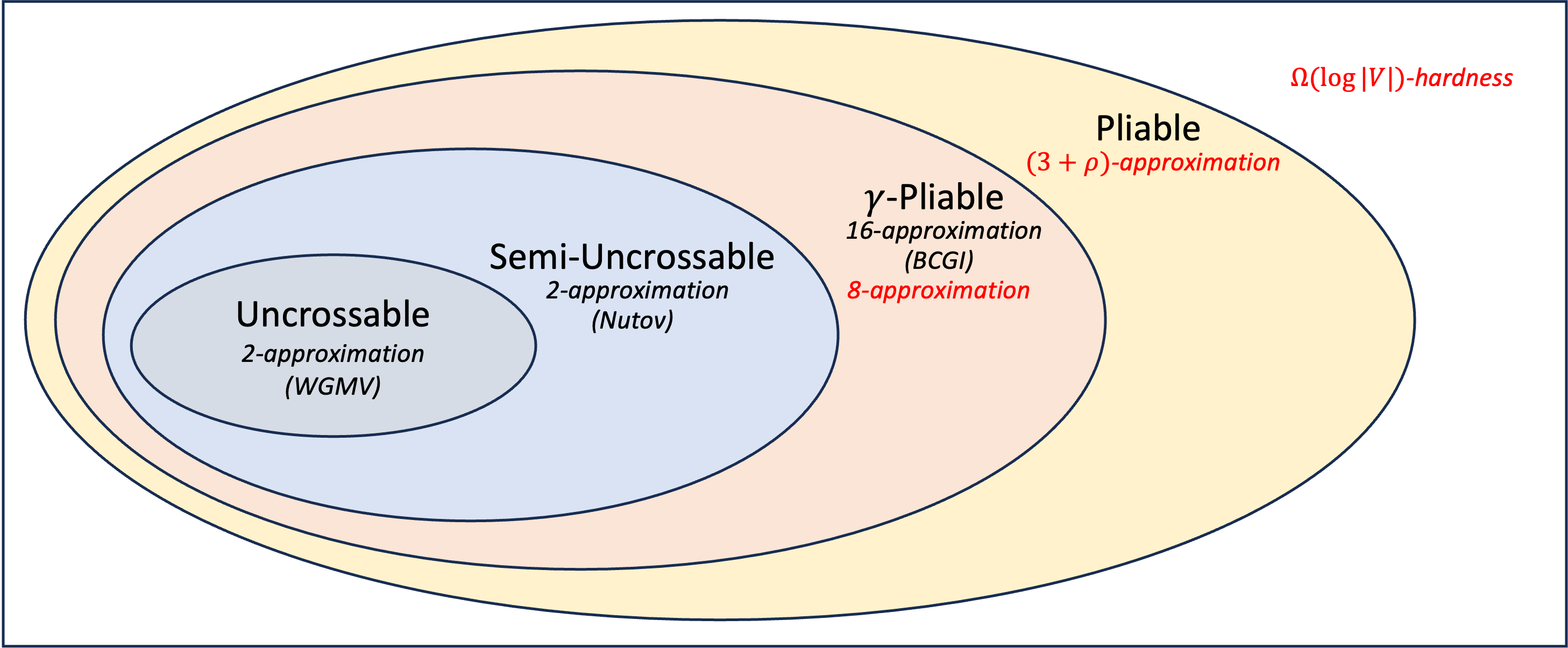}
    \caption{Performance of the WGMV primal-dual algorithm for the $\mathcal{F}$-augmentation problem. Red text indicates work in this paper.}
    \label{fig:StateOfArt}
\end{figure}

Prior to our work however, nothing was known about the performance of the WGMV primal-dual algorithm for general pliable families. We bridge this gap by introducing a novel concept known as the crossing density of a set family that captures a notion of how often sets within the family cross\footnote{Recall that two sets $A$ and $B$ cross if all four sets $A\setminus B, B\setminus A, A\cap B, \overline{A\cup B}$ are non-empty.}. We show that the WGMV primal-dual algorithm for the $\mathcal{F}$-augmentation problem when $\mathcal{F}$ is a general pliable family provides an approximation ratio parametrized by the crossing density of the family. Furthermore, this performance is nearly asymptotically tight and cannot be improved much. In addition, we show that attempts to generalize pliable families further by requiring only one of the four corner cuts to lie within the family are futile, as the $\mathcal{F}$-augmentation problem then captures the hitting set problem and has an $\Omega(\log |V|)$-hardness of approximation (See Figure \ref{fig:StateOfArt}).

As detailed in the next section, we also apply our techniques to specific network design problems to obtain current best approximation ratios. In particular, we study the small cut augmentation problem, the capacitated $k$-edge connectivity problem and the flexible graph connectivity problem.

\subsection{Our Contribution, Techniques and Comparison with Previous Work}

\subsubsection{Crossing Density}~

Our first major contribution is the identification of the right parameter to study when analyzing the WGMV primal-dual algorithm for non-uncrossable families. To quote WGMV, ``The key feature of uncrossable functions is that there exists an optimal dual solution that is laminar''. WGMV heavily relied on this \textit{dual laminarity} property in the design of their algorithm but BCGI observed that this property does not hold for pliable families. The main obstacle seems to be that inclusion-wise minimal sets of the family $\mathcal{F}$ (also referred to as \textit{active sets}) start to cross other sets within the family. It is simple to see that such crossings do not occur in the context of uncrossable families. Indeed if $C$ is an active set that crosses $S\in\mathcal{F}$, then the uncrossable property will imply that either $C\cap S$ or $C\setminus S$ is also in $\mathcal{F}$ contradicting the minimality of the active set $C$. BCGI attempted to control for such undesirable crossings by considering a sub-family of pliable families characterized by a set property they called \textbf{property $\bm{\gamma}$}. For any set of edges $D\subseteq E$, define $\mathcal{F}_D := \{S\in\mathcal{F}:\delta(S)\cap D = \emptyset\}$ i.e. the family of sets in $\mathcal{F}$ that have not yet been augmented by edges in $D$. Define $\mathcal{C}_D$ to be the inclusion-wise minimal sets in $\mathcal{F}_D$.

\begin{definition}[$\gamma$-Pliable Family]\label{def:gammapliable}
    A set family $\mathcal{F}$ is ${\gamma}${-pliable} if it is pliable and also satisfies \\
\textbf{Property }$\bm{\gamma}$: For any edge set $D\subseteq E$, sets $S_1 \subseteq S_2 \in \mathcal{F}_D$, and $C \in \mathcal{C}_D$ that crosses both $S_1$ and $S_2$, the set $S_2 \setminus (S_1\cup C)$ is either empty or lies in $\mathcal{F}$.
\end{definition}

BCGI utilized this local property and showed how to extend the WGMV primal-dual method beyond uncrossable families. As evidenced by the 30 year hiatus on progress along these lines, BCGI had to introduce novel analysis techniques for the primal-dual algorithm. However, they were unable to capture pliable families in general because their methodology for handling crossings between active sets and other sets within the family $\mathcal{F}$ was local. Instead, we explicitly and globally account for such crossings by defining the crossing density of a set family.

\begin{definition}[Crossing Density]\label{def:CrossingDensity}
    The crossing density of a set family $\mathcal{F}$ is the smallest number $\rho$ such that for any edge set $D\subseteq E$ and a laminar subfamily $\mathcal{F}_D^{lam}$ of $\mathcal{F}_D$, we have
\[ \text{number of crossing pairs $(S,C)$ where $S\in\mathcal{F}_D^{lam}$ and $C\in\mathcal{C}_D$ }
\leq \rho |\mathcal{C}_D|
\]
\end{definition}

The definition of crossing density above compares the number of crossings between active sets and other sets within the family with the total number of active sets. The smaller the density, the more `manageable' a set family is. Note that the crossing density of uncrossable families is zero since active sets do not cross other sets. The reason we restrict our attention to only laminar sub-families $\mathcal{F}_D^{lam}$ of $\mathcal{F}_D$ is due to the analysis technique of the primal-dual algorithm we utilize that relies on the construction of laminar sub-families of $\mathcal{F}_D$. If we consider all sets $S \in \mathcal{F}_D$, then the crossing density could be made artificially larger. To the best of our knowledge, a notion similar to the crossing density has not been studied in the past in the context of network design. We are able to prove the following theorem using this novel idea.

\begin{theorem}\label{thm:crossdensity}
    The WGMV primal-dual algorithm is a $(3+\rho)$-approximation algorithm for the $\mathcal{F}$-augmentation problem when $\mathcal{F}$ is a pliable family with crossing density $\rho$.
\end{theorem}

We supplement this theorem by showing that it is nearly asymptotically tight. There exists a family of instances of the $\mathcal{F}$-augmentation problem on pliable families where the crossing density is $O(|V|)$ and the WGMV primal-dual algorithm provides an approximation ratio of $\Omega(\sqrt{|V|})$. Hence, the approximation factor in Theorem \ref{thm:crossdensity} can only be improved to $O(\sqrt{\rho})$ and not further. We provide details of this example in Appendix \ref{apx:tightresult}. Furthermore, we show that one cannot hope to extend pliable families further in the most natural way. Suppose we relax the pliability condition and require that just one of the corner cuts (instead of two) also lie within the family, then the problem becomes as hard as the hitting set problem ($\Omega(\log |V|)$-hardness of approximation). We provide the details of this reduction in Appendix \ref{apx:beyondpliable}.

The techniques we use to prove Theorem \ref{thm:crossdensity} rely on the age-old idea of constructing a laminar witness set family for the output set of edges $H$ from the primal-dual algorithm, and devising a clever token distribution scheme to bound the average degree of active sets with respect to $H$ i.e. $|\delta(C)\cap H|$. This methodology has been frequently used when analyzing primal-dual techniques \cite{williamson1993primal,bansal2024improved,ravi1997approximation,vegh2008primal,melkonian2005primal,friggstad2023constant} and the novelty often lies in the construction of the laminar witness family and/or in the token distribution scheme. Fortunately, BCGI had already constructed a laminar witness family in the context of pliable families and we can borrow this idea from them. However, most token distribution schemes in the literature are local and rely on local properties like uncrossability and property $\gamma$. We introduce a novel and more global token distribution scheme that utilizes the crossing density to obtain a bound on the average degree of active sets. The details are deferred to Section \ref{sec:crossdensity}. We hope that our methodology and the idea of crossing density will be applicable to other algorithmic techniques that rely on dual laminarity as well, like iterative rounding for instance.

\begin{remark}
    BCGI showed a 16-approximation factor for $\gamma$-pliable families while we show an $O(1)$-approximation algorithm when the crossing density is $O(1)$. One might think that the two concepts (bounded crossing density and property $\gamma$) are related but this is not the case. We show examples of set families in Appendix \ref{appendix:crossversusgamma} where one holds and not the other, and vice versa.
\end{remark}

\subsubsection{Small Cut Augmentation}~

Our second major contribution is in the context of $\alpha$-min cuts. One of the major motivations for considering pliable families (apart from it being a natural generalization of uncrossable families) seems to be that it captures the family of $\alpha$-min cuts. The problem of augmenting min cuts has been extensively studied under the name of weighted connectivity augmentation and there has been a steady line of work leading to sustained progress towards the problem. The problem of augmenting $\alpha$-min cuts however has proven to be more involved since the family is not uncrossable. BCGI introduced the \textbf{Small Cut Augmentation} problem to capture the problem of augmenting $\alpha$-min cuts.

\begin{problem}[Small Cut Augmentation] ~\\ \textit{Input:} Graph $G=(V,E\cup E_0)$ with edge costs $\{c_e\}_{e\in E}$, edge capacities $\{u_e\}_{e\in E_0}$ and a threshold $k>0$. \\ \textit{Output:} Subset of the edges $F\subseteq E$ with minimum $c(F) = \sum_{e\in F}c_e$ such that for every non-empty $S\subsetneq V$ with capacity $\sum_{e\in \delta(S)\cap E_0}u_e < k$, the set of edges $F\cap\delta(S)$ is non-empty.  
\end{problem}

The family $\mathcal{F}$ of cuts $S$ with capacity $\sum_{e\in \delta(S)\cap E_0}u_e < k$ are referred to as \textit{small cuts} and it is easy to see that changing the threshold $k$ captures the family of $\alpha$-min cuts for any $\alpha>0$. BCGI showed that the family of small cuts is a $\gamma$-pliable family and so were able to provide a 16-approximation algorithm for the small cut augmentation problem\footnote{Nutov \cite{nutov2024improved} improved this to a 10-approximation algorithm recently}. We have already seen that the family of small cuts $\mathcal{F}$ is pliable. To observe that property $\gamma$ holds, we go back to the submodularity \eqref{eq:submodularity} and posimodularity \eqref{eq:posimodularity} of the cut function. Let $C \in \mathcal{C}_D$ that crosses both $S_1\subseteq S_2 \in \mathcal{F}_D$. Then, equation \eqref{eq:submodularity} tells us that one of $S_1\cup C$ or $S_1\cap C$ lies in $\mathcal{F}$. However, $C$ is an inclusion-wise minimal set and so $S_1\cup C \in \mathcal{F}$. Now if $S_2\setminus (S_1\cup C)$ is non-empty, then $S_2$ crosses $S_1\cup C$. Equation \eqref{eq:posimodularity} tells us that either $(S_1\cup C)\setminus S_2$ or $S_2\setminus (S_1\cup C)$ lies in $\mathcal{F}$. But again, $(S_1\cup C)\setminus S_2$ is a proper subset of $C$ and so $S_2\setminus (S_1\cup C)$ lies in $\mathcal{F}$ proving property $\gamma$.

To improve on previous approximation factors for the small cut augmentation problem, we make a critical observation about the family of small cuts i.e. they satisfy a particular sparse crossing property defined below.

\begin{definition}[Sparse Crossing Property]\label{def:sparsecrossing}
    A set family $\mathcal{F}$ has the sparse crossing property if for any subset of the edges $D\subseteq E$, a set $S\in\mathcal{F}_D$ crosses at most one set in $\mathcal{C}_D$.
\end{definition}

We extend the analysis in BCGI \cite{bansal2024improved} by using the above sparse crossing property and prove the following theorem.

\begin{theorem}\label{thm:6-appx}
    There exists a 6-approximation algorithm for the $\mathcal{F}$-augmentation problem when $\mathcal{F}$ is a $\gamma$-pliable family satisfying the sparse crossing property.
\end{theorem}

By showing that the family of small cuts satisfies the sparse crossing property, we obtain the following theorem as a corollary.


\begin{theorem}\label{thm:SmallCuts}
    There exists a 6-approximation algorithm for the Small Cut Augmentation Problem.
\end{theorem}

BCGI also showed how an algorithm for the small cut augmentation problem can be used to solve the capacitated $k$-edge connected spanning subgraph problem. Here one is given a graph Graph $G=(V,E)$ with edge costs $\{c_e\}_{e\in E}$, edge capacities $\{u_e\}_{e\in E}$ and the goal is to find a cheapest subgraph so that the min cut in the graph has capacity at least $k$. Using Theorem \ref{thm:SmallCuts}, we can improve on previous approximation ratios for this problem to obtain a $6\lceil k/u_{min}\rceil$-approximation algorithm.

\subsubsection{Flexible Graph Connectivity}~

Our final major contribution is in the study of the flexible graph connectivity (FGC) problem. This very interesting network design model was introduced by Adjiashvili et al. \cite{adjiashvili2022flexible}. It is among the non-uniform fault tolerant models that have become increasingly popular within the network design community \cite{adjiashvili2022flexible,boyd2024approximation,bansal2024improved,bansal2022algorithms,chekuri2023,nutov2023improved,hyattdenesik2024improved}. Boyd et al. \cite{boyd2024approximation} generalized this model to the $(p,q)$-FGC problem.

\begin{problem}[~$(p,q)$-Flexible Graph Connectivity] ~\\ \textit{Input:} Graph $G=(V,E)$ with edge costs $\{c_e\}_{e\in E}$, and a partition of the edge set into safe edges $\mathcal{S}$ and unsafe edges $\mathcal{U}$.\\ \textit{Output:} Subset of the edges $F\subseteq E$ with minimum $c(F) = \sum_{e\in F}c_e$ such that $(V,F)$ is $p$-edge connected even after the removal of any $q$ unsafe edges from $F$.
\end{problem}

A popular conjecture for the FGC problem is that when $q$ is small, constant factor approximation algorithms should be obtainable \cite{chekuri2023}. However, little progress has been made towards this conjecture. $(p,0)$-FGC is simply the $p$-edge connected spanning subgraph problem and 2-approximation algorithms have been known for a while. $(p,1)$-FGC approximately reduces to the augmentation problem of uncrossable families \cite{boyd2024approximation} and so an $O(1)$-approximation algorithm is obtainable. $(p,2)$-FGC approximately reduces to the augmentation problem of $\gamma$-pliable families \cite{bansal2024improved} and so an $O(1)$-approximation algorithm is obtainable here as well. However, as observed by BCGI \cite{bansal2024improved} and Chekuri, Jain \cite{chekuri2023}, the $(p,3)$-FGC problem does not approximately reduce to uncrossable or pliable augmentation problems and so no constant factor approximation was known prior to our work. The previous best approximation ratio for this problem is $O(p)$ \cite{chekuri2023}. We prove the following theorem.

\begin{theorem}\label{thm:FGC}
    $(p,3)$-FGC admits a 12-approximation algorithm when $p$ is even and an $11+\epsilon$-approximation algorithm when $p$ is odd.
\end{theorem}

Our techniques here are different depending on the parity of $p$. When $p$ is even, we start with a $6$-approximate solution to the $(p,2)$-FGC problem \cite{bansal2024improved} and attempt to augment the family of cuts that are still violated - a cut is violated if it has fewer that $p$ safe edges and fewer than $p+3$ edges in total. We show that the family of violated cuts forms a pliable family with the sparse crossing property. Thus Theorem \ref{thm:6-appx} implies that we can find a 6-approximate solution to the resulting augmentation problem. Putting it all together, we obtain an 12-approximation algorithm. 

When $p$ is odd, we start with a $7+\epsilon$-approximate solution to the $(p,2)$-FGC problem \cite{nutov2023improved}. Unfortunately the resulting family of violated cuts is not pliable anymore. Instead, we painstakingly split these cuts into two subfamilies based on how they cross each other. To do so, we introduce the novel idea of a set $A$ \textit{amenably crossing} a set $B$ based on the existence of unsafe edges with one end point in $A\cap B$ and the other in $B\setminus A$. By developing useful tools to analyze crossings between three sets we show that the subfamily of all violated sets that amenably cross some other violated set forms an uncrossable family. Interestingly, the subfamily of all other violates sets is a laminar family and so is also uncrossable. Since the definition of amenable crossings is local, we can find these subfamilies in polynomial time, and augment them separately using the WGMV primal-dual algorithm. Overall, we obtain an $11+\epsilon$-approximation algorithm. The details are deferred to Section \ref{sec:FGC}. To the best of our knowledge, the idea of splitting families of cuts based on how the sets cross each other has not been previously seen in the literature. In addition, the tools we develop to analyze crossings between multiple sets might be useful for other FGC problems with larger values of $q$.

\subsubsection{Improvements for $\gamma$-Pliable Families}~

As a byproduct of our analysis, we also improve the results by BCGI \cite{bansal2024improved} and Nutov \cite{nutov2024improved} on the $\mathcal{F}$-augmentation problem when $\mathcal{F}$ is a $\gamma$-pliable family. BCGI had shown a 16-factor approximation and Nutov tightened their analysis to obtain a 10-approximation. As observed by BCGI, these improvements immediately help obtain better approximation ratios for the Capacitated $k$-edge connected spanning subgraph problem and the small cut augmentation problem. We go further, improving the analysis to obtain an 8-approximation.

\begin{theorem}\label{thm:8appx}
    There exists an 8-approximation algorithm for the $\mathcal{F}$-augmentation problem when $\mathcal{F}$ is a $\gamma$-pliable family.
\end{theorem}

\begin{table}[H]
    \centering
    \begin{tabular}{|c|c|c|}
    \hline \textbf{Problem} & \textbf{previous} & \textbf{this paper} \\
      \hline \hline $\mathcal{F}$-augmentation of pliable families   & --- & $3+\rho$\\
       \hline $\mathcal{F}$-augmentation of $\gamma$-pliable families  & 10 \cite{nutov2024improved} & 8 \\
        \hline Small Cut Augmentation & 10 \cite{nutov2024improved} & 6\\
        \hline Capacitated $k$-ECSS & $10\cdot \lceil k/u_{min}\rceil$ \cite{nutov2024improved} & $6\cdot \lceil k/u_{min}\rceil$\\
        \hline $(p,3)$-FGC & $O(p)$ \cite{chekuri2023} & 12\\
        \hline
    \end{tabular}
    \caption{Summary of results in this paper. $\rho$ is the crossing density of a set family and $u_{min}$ is the minimum capacity of an edge}
    \label{tbl:results}
\end{table}
\vspace{-10mm}
\subsection{Related Work}

The $\mathcal{F}$-augmentation problem was introduced by WGMV \cite{williamson1993primal} in an attempt to study the more general $f$-connectivity problem where each cut $S$ has a required number of edges to be bought from $\delta(S)$. They provided a primal dual 2-approximation algorithm for $\mathcal{F}$-augmentation when $\mathcal{F}$ is uncrossable and used this to study the $f$-connectivity problem when $f$ is a so-called \textit{proper} function. Bansal et al. \cite{bansal2024improved} studied the $\mathcal{F}$-augmentation problem beyond uncrossable families and introduced pliable families. They provided a 16-approximation algorithm for $\mathcal{F}$-augmentation of $\gamma$-pliable families. Their analysis was later refined by Nutov \cite{nutov2024improved} who showed a 10-approximation factor. Nutov \cite{nutov2023extend} studied \textit{semi-uncrossable} families which are a subclass of pliable families but more general than uncrossable families and showed that the primal-dual algorithm is a 2-approximation algorithm here. Jain \cite{Jain01} introduced a different approach to solving augmentation and connectivity problems known as iterative rounding. He provided a 2-approximation algorithm for a variety of network design problems including the survivable network design problem. We should point out that the $\mathcal{F}$-augmentation problem has been studied in very specific settings as well like tree augmentation, forest augmentation, and connectivity augmentation. The current goal here is to obtain approximation algorithms with smaller constant factors as both WGMV \cite{williamson1993primal} and Jain \cite{Jain01} immediately provide a 2-approximation algorithm for these problems. We refer the reader to works by Traub et al. and Ravi et al. for recent developments \cite{traub2022local,traub2023,ravi2023approximation}.

The capacitated $k$-edge connected spanning subgraph problem was introduced by Goemans et al. \cite{goemans1994improved} who showed that there is a $2k$-approximation algorithm for the problem. Carr et al. \cite{carr1999strengthening} introduced the very versatile knapsack cover inequalities to strengthen the LP relaxation of the problem and exhibited an LP rounding scheme to provide an approximation factor characterized by the size of the largest \textit{bond} in the graph. Chakrabarty et al. \cite{chakrabarty2015approximability} used these knapsack cover inequalities to provide an $O(\log |V|)$-approximation algorithm based on randomized rounding. Boyd et al. \cite{boyd2024approximation} exhibited a $k$-approximation and Bansal et al. \cite{bansal2024improved} provided a $16\cdot\lceil k/u_{min}\rceil$ approximation. This was later improved by Nutov \cite{nutov2024improved} to $10\cdot\lceil k/u_{min}\rceil$.

The small cut augmentation problem was introduced by Bansal et al. \cite{bansal2024improved} who provided a 16-approximation algorithm for the problem. This was improved to a factor of 10 by Nutov \cite{nutov2024improved}. Nutov \cite{nutov2023improved} used a repeated augmentation framework to provide a $(\alpha - \lambda_0)$-approximation algorithm where $\alpha$ is the threshold defining the small cuts and $\lambda_0$ is the size of the mincut in the graph. The structure of near minimum cuts has also been studied by Karger \cite{Karger93}, Karger and Stein \cite{karger1996new}, and Benczur and Goemans \cite{benczur2008deformable}.

The flexible graph connectivity problem was first studied in the shortest path setting \cite{adjiashvili2013fault}. The model we study was introduced by Adjiashvili et al. \cite{adjiashvili2022flexible} and generalized by Boyd et al. \cite{boyd2024approximation} to the $(p,q)$-FGC model. Adjiashvili et al. \cite{adjiashvili2022flexible} presented the first constant factor approximation for $(1,1)$-FGC. This was improved to a factor of 2 by Boyd et al. \cite{boyd2024approximation} who also provided a $(q+1)$-approximation for $(1,q)$-FGC, a 4-approximation for $(p,1)$-FGC and an $O(q\log n)$-approximation for the general $(p,q)$-FGC problem. Chekuri and Jain \cite{chekuri2023} presented a $(2q+2)$-approximation algorithm for $(2,q)$-FGC and also an $O(p)$-approximation for $(p,2)$-FGC, $(p,3)$-FGC and $(2p,4)$-FGC. Bansal et al. \cite{bansal2024improved} provided the first constant factor approximation for $(p,2)$-FGC exhibiting a 6-approximation when $p$ is even and a $20$-approximation when $p$ is odd. This was later improved by Nutov \cite{nutov2023improved} to a $(7+\epsilon)$-approximation when $p$ is odd. Hyatt-Denesik et al. \cite{hyattdenesik2024improved} study an unweighted version of $(1,q)$-FGC and exhibit a $1+O(1/\sqrt{q})$-approximation algorithm in this case. Bansal et al. \cite{bansal2022algorithms} study a Steiner version of $(1,1)$-FGC from a parameterized algorithms lens.

\section{Preliminaries}\label{sec:prelims}

We begin by re-explaining the WGMV \cite{williamson1993primal} primal-dual algorithm at a high level. The WGMV algorithm is based on the following linear program relaxation for the $\mathcal{F}$-augmentation problem and its dual.

\noindent
\begin{minipage}[t]{0.45\textwidth} \label{eq:primaldualLPs}
\footnotesize
\begin{center}
    \textbf{Primal LP}
\end{center}
\vspace{-15pt}
\begin{align}
    & \qquad \, \, \, \min \quad \sum_{e \in E} c_e x_e & \notag \\
    & \text{subject to:} \sum_{e \in \delta(S)} x_e \geq 1 \quad \forall S \in \mathcal{F} \notag \\
    & \qquad \qquad \quad  0 \leq x_e \leq 1 \quad \, \, \, \forall e \in E \notag
\end{align}
\end{minipage}
\quad \vline \quad 
\begin{minipage}[t]{0.45\textwidth}
\footnotesize
\begin{center}
    \textbf{Dual LP}
\end{center}
\vspace{-15pt}
\begin{align}
& \qquad \, \, \max \quad \, \, \, \sum_{S \in \mathcal{F}} y_S \notag \\
& \text{subject to:} \sum_{S \in \mathcal{F} : e \in \delta(S)} y_S \leq c_e \quad \forall e \in E \notag \\
& \qquad \qquad \qquad y_S \geq 0 \qquad \qquad \, \, \, \forall S \in \mathcal{F} \notag
\end{align}
\end{minipage}

\medskip

The algorithm starts with an infeasible primal solution $F =
\emptyset$, which corresponds to $x = \chi^F = \mathbf{0}$, and a feasible dual solution $y = \mathbf{0}$. At any time, we refer to the sets in $\mathcal{F}_F = \{S\in\mathcal{F} : \delta(S)\cap F = \emptyset\}$ as \textit{violated sets} and its inclusion-wise minimal elements as \textit{active sets}. We will denote the family of active sets using $\mathcal{C}$. The algorithm has two stages. In the first stage, the algorithm
iteratively increases the dual values of all active sets uniformly till an edge $e\in E\setminus F$ becomes tight, and adds edge $e$ to the set $F$. The first stage ends when $\mathcal{F}_F = \emptyset$. In the second stage, the algorithm removes redundant edges in reverse order of additions. This final set of edges $F$ is returned by the algorithm. WGMV showed the following theorem.

\begin{theorem}[\cite{williamson1993primal}]\label{thm:WGMV}
    Let $F$ be the set of edges returned by the primal-dual algorithm. If in every iteration of the first stage of the primal-dual algorithm,
\[
\sum_{C\in \mathcal{C}}\left|\delta(C)\cap F\right| \leq \alpha \left|\mathcal{C}\right|
\]
where $\mathcal{C}$ is the set of active sets in that iteration, then the algorithm is an $\alpha$ approximation-algorithm for the $\mathcal{F}$-augmentation problem.
\end{theorem}

They showed that $\alpha = 2$ is valid when $\mathcal{F}$ is uncrossable and Bansal et al. \cite{bansal2024improved} showed that $\alpha = 16$ is valid when $\mathcal{F}$ is $\gamma$-pliable. A note on oracle access required by the WGMV algorithm and future works on the primal-dual algorithm: For any subset of the edges $D\subseteq E$, the algorithm needs oracle access to the inclusion-wise minimal sets from the set family $\mathcal{F}_D$. We will assume throughout this paper that such an oracle exists. However, when applying the WGMV algorithm to specific network design problems, the existence of such an oracle has to be proven (see \cite{williamson1993primal,bansal2024improved} for example).

\subsection{Witness Sets and their Red-Green Coloring}

We mention some lemmas from \cite{bansal2024improved} that will be useful to us. We use the shorthand $\delta_F(C)$ to mean $\delta(C)\cap F$.

\begin{lemma}[\cite{bansal2024improved}]\label{lem:laminarwitness}
    Suppose $\mathcal{F}$ is a pliable family. For any edge $e\in \cup_{C\in\mathcal{C}}\delta_F(C)$, there exists a witness set $S_e \in \mathcal{F}$ such that (i) $S_e$ is violated in the current iteration and (ii) $\delta_F(S_e) = \{e\}$.\\
    Furthermore, there exists a laminar family of witness sets for edges in $\cup_{C\in\mathcal{C}}\delta_F(C)$.
\end{lemma}

An analogous lemma was proven by Williamson et al. \cite{williamson1993primal} in the context of uncrossable families and Bansal et al. \cite{bansal2024improved} generalized this lemma in the context of pliable families.

Let $\mathcal{L}$ be the laminar family of the witness sets from Lemma \ref{lem:laminarwitness} together with the node set $V$. This laminar family can be represented by a rooted directed out-tree $\mathcal{T}$ with a node $v_S$ for every set $S\in\mathcal{L}$. The root is the node $v_V$ and there is an edge from $v_S$ to $v_T$ if and only if $S$ is the smallest set in $\mathcal{L}$ that properly contains $T$. A node $v_S$ in $\mathcal{T}$ is called active if there exists an active set $C\in\mathcal{C}$ such that $S$ is the smallest set in $\mathcal{L}$ that contains $C$. We color all active nodes using \textbf{red}, and color all non-red nodes with out degree at least 2 using \textbf{green}. Every other node is colored black (uncolored). Let $n_R, n_G$ and $n_B$ be the number of red, green and black nodes respectively. Let $n_L$ be the number of leaves in $\mathcal{T}$

\begin{lemma}[\cite{bansal2024improved}]\label{lem:colornodes}The following statements are true.
    \begin{enumerate}
        \item[(i)] The sets in $\mathcal{C}$ are pairwise disjoint.
        \item[(ii)] Every leaf node of $\mathcal{T}$ is colored red.
        \item[(iii)] $n_G\leq n_L \leq n_R \leq \left|\mathcal{C}\right|$. 
    \end{enumerate}
\end{lemma}

A proof of this lemma is given in Appendix \ref{appendix:lemproof}. We end this section with a simple observation about crossings between active sets and witness sets.

\begin{lemma}\label{obs1}
    Let $C\in\mathcal{C}$ and $S\in \mathcal{L}\setminus\{V\}$. If $S\cap C$ and $C\setminus S$ are non-empty then $S$ crosses $C$.
\end{lemma}
\begin{proof}
    If $S\setminus C$ is empty, then $S\subsetneq C$ contradicting the minimality of $C$. If $\overline{S\cup C}$ is empty, then one of $S\cap C$ or $C\setminus S$ must also be violated due to pliable property. But then again, a proper subset of $C$ is violated which is a contradiction. \qed
\end{proof}

\section{Crossing Density}\label{sec:crossdensity}

In this section, we prove Theorem \ref{thm:crossdensity} exhibiting an approximation algorithm for the $\mathcal{F}$-augmentation of pliable families based on the crossing density of the family $\mathcal{F}$. We start by distributing some tokens to $\mathcal{T}$ and make some observations and claims.\\

\noindent\fbox{\parbox{\textwidth}{\textit{Token Distribution Phase I}\\ For every active set $C\in\mathcal{C}$ and $e\in \delta_F(C)$, assign a token to the node $v_{S_e}$ where $S_e$ is the witness set such that $\delta_F(S_e) = \{e\}$.}}

~\\Observe that the number of tokens distributed is exactly equal to $\sum_{C\in\mathcal{C}}\delta_F(C)$. Furthermore, the node $v_{S_e}$ has tokens if there are active sets $C$ with $e\in \delta(C)$. There can be at most two such active sets each containing an end point of edge $e$ exclusively (since active sets are disjoint by Lemma \ref{lem:colornodes}). Thus, each node in $\mathcal{T}$ has at most two tokens. Additionally, each non-root node $v_S$ in $\mathcal{T}$ has at least one token since $S$ is a witness set and so $\delta_F(S) = \{e\}$ for some edge $e\in \cup_{C\in\mathcal{C}}\delta_F(C)$. Note that the root node $v_V$ does not have tokens and will be irrelevant to our discussions henceforth.

\begin{claims}\label{clm:longchain}
    Let $v_{S_1},v_{S_2},\ldots, v_{S_k}$ be a directed path in $\mathcal{T}$ where none of $v_{S_1},v_{S_2},\ldots, v_{S_{k-1}}$ is red. Let $e_k = (u,v)$ be the edge such that $\delta_F(S_k) = \{e_k\}$ with $u\in S_k$. Then,
    \begin{enumerate}
        \item[(i)] If there exists $C\in\mathcal{C}$ with $v\in C, u\not\in C$, then $C$ crosses all the sets ${S_1},{S_2},\ldots, {S_{k-1}}$.
        \item[(ii)] If there exists $C\in\mathcal{C}$ with $u\in C, v\not\in C$ and $v_{S_k}$ is not red, then $C$ crosses all the sets ${S_1},{S_2},\ldots, {S_{k}}$.
    \end{enumerate}
\end{claims}

\begin{proof}
    \begin{enumerate}
        \item[(i)] The active set $C$ cannot be contained in ${S_k}$ since $v\in C\setminus S_k$. Also $C$ is not contained in any of $S_{1},\ldots,S_{k-1}$, as otherwise one of these sets must be the smallest witness set containing $C$ and the corresponding node $v_S$ will be colored red. Furthermore, $v\in S_i$ for every $i\leq k-1$ since otherwise $e_k\in\delta(S_i)$ contradicting property (ii) of witness sets in Lemma \ref{lem:laminarwitness}. $C\setminus S_i$ are non-empty and so $C$ crosses $S_i$ (by Lemma \ref{obs1}).

        \item[(ii)] The active set $C$ cannot be contained in $S_k$ since otherwise there would have to exist a witness set $S$ such that $C\subseteq S \subsetneq S_k$. But then $e_k\in\delta(S)$ contradicting property (ii) of witness sets in Lemma \ref{lem:laminarwitness}. As before, this implies that $C$ is not contained in any of $S_{1},\ldots,S_{k-1}$. But now for all $i\leq k$, $C\cap S_i$ is non-empty as $u\in C\cap S_i$ and $C\setminus S_i$ is non-empty. Hence $C$ crosses all the sets $S_1,\ldots,S_k$.\qed
    \end{enumerate}
\end{proof}
{
\begin{figure}[htb] \centering
{
\begin{tikzpicture}[scale=0.8]
\begin{scope}[every node/.style={circle, fill=black, draw, inner sep=0pt, minimum size = 0.15cm, label distance=0.1cm}]
	\node[draw=none,fill=none]	(v04) at ($(0,0)+(-6,1.5)$) {};
	\node[draw=none,fill=none]	(v03) at ($(0,0)+(-5,1.5)$) {};
	\node[draw=none,fill=none]	(v02) at ($(0,0)+(-4,1.5)$) {};
	\node[draw=none,fill=none]	(v01) at ($(0,0)+(-3,1.5)$) {};
	\node[draw=none,fill=none]			(v1) at ($(0,0)+(1,1.5)$) {};
	\node[draw=none,fill=none]			(v2) at ($(0,0)+(2,1.5)$) {};
	\node[draw=none,fill=none]			(v3) at ($(0,0)+(3,1.5)$) {};
	\node[draw=none,fill=none]			(v4) at ($(0,0)+(4,1.5)$) {};
	\node[draw=none,fill=none,label={[label distance=0.0]0:$S_4$}] (s4) at ($(0,0)+(-2.25,1.5)$) {};
	\node[label={[label distance=0.0]180:$u$}]	(u) at ($(0,0)+(-0.5,1.5)$) {};
	\node[label={[label distance=0.0]270:$v$}]	(v) at ($(0,0)+(+0.8,1.5)$) {};
	\node[draw=none,fill=none,label={[label distance=0.0]0:$S_3$}] (s3) at ($(0,0)+(-3.00,1.5)$) {};
	\node[draw=none,fill=none,label={[label distance=0.0]0:$S_2$}] (s2) at ($(0,0)+(-4.00,1.5)$) {};
	\node[draw=none,fill=none,label={[label distance=0.0]0:$S_1$}] (s1) at ($(0,0)+(-5.00,1.5)$) {};
\end{scope}

\begin{scope}[line width=0.6pt]
	\draw[thick,dashed] ($0.5*(v01)+0.5*(v1)$) ellipse (1.25cm and 0.60cm);

	\draw[thick,dashed] ($0.5*(v02)+0.5*(v2)$) ellipse (2.00cm and 1.00cm);

	\draw[thick,dashed] ($0.5*(v03)+0.5*(v3)$) ellipse (3.00cm and 1.50cm);

	\draw[thick,dashed] ($0.5*(v04)+0.5*(v4)$) ellipse (4.00cm and 2.00cm);
\end{scope}

        \begin{scope}[every edge/.style={draw=black}]
                                
                \path[thick] (u) edge[] node {} (v);  
        \end{scope}
\draw[thick] (2,1.5) circle (1.5); \node at (3.8,1.5) {$C$};

\begin{scope}[every node/.style={circle, fill=black, draw, inner sep=0pt, minimum size = 0.15cm, label distance=0.1cm}, shift = {(10, 0)}]
	\node[draw=none,fill=none]	(v04x) at ($(0,0)+(-6,1.5)$) {};
	\node[draw=none,fill=none]	(v03x) at ($(0,0)+(-5,1.5)$) {};
	\node[draw=none,fill=none]	(v02x) at ($(0,0)+(-4,1.5)$) {};
	\node[draw=none,fill=none]	(v01x) at ($(0,0)+(-3,1.5)$) {};
	\node[draw=none,fill=none]			(v1x) at ($(0,0)+(1,1.5)$) {};
	\node[draw=none,fill=none]			(v2x) at ($(0,0)+(2,1.5)$) {};
	\node[draw=none,fill=none]			(v3x) at ($(0,0)+(3,1.5)$) {};
	\node[draw=none,fill=none]			(v4x) at ($(0,0)+(4,1.5)$) {};
	\node[draw=none,fill=none,label={[label distance=0.0]0:$S_4$}] (s4x) at ($(0,0)+(-2.25,1.5)$) {};
	\node[label={[label distance=0.0]180:$u$}]	(ux) at ($(0,0)+(-0.5,1.5)$) {};
	\node[label={[label distance=0.0]270:$v$}]	(vx) at ($(0,0)+(+0.8,1.5)$) {};
	\node[draw=none,fill=none,label={[label distance=0.0]0:$S_3$}] (s3x) at ($(0,0)+(-3.00,1.5)$) {};
	\node[draw=none,fill=none,label={[label distance=0.0]0:$S_2$}] (s2x) at ($(0,0)+(-4.00,1.5)$) {};
	\node[draw=none,fill=none,label={[label distance=0.0]0:$S_1$}] (s1x) at ($(0,0)+(-5.00,1.5)$) {};
\end{scope}

\begin{scope}[line width=0.6pt, shift = {(10, 0)}]
	\draw[thick,dashed] ($0.5*(v01x)+0.5*(v1x)$) ellipse (1.25cm and 0.60cm);

	\draw[thick,dashed] ($0.5*(v02x)+0.5*(v2x)$) ellipse (2.00cm and 1.00cm);

	\draw[thick,dashed] ($0.5*(v03x)+0.5*(v3x)$) ellipse (3.00cm and 1.50cm);

	\draw[thick,dashed] ($0.5*(v04x)+0.5*(v4x)$) ellipse (4.00cm and 2.00cm);
\end{scope}

        \begin{scope}[every edge/.style={draw=black}, shift = {(10, 0)}]
                                
                \path[thick] (ux) edge[] node {} (vx);  
        \end{scope}
\draw[thick] (7.3,1.5) circle (2.5); \node at (4.8,2.5) {$C$};

\end{tikzpicture}
}
\caption{
        \label{fig:Claim1}
	Illustration of Claim 1. If $v\in C$, then $C$ crosses the sets $S_1,S_2,S_3$. If $u\in C$, then $C$ crosses all the sets $S_1,\ldots, S_4$. Otherwise one of the corresponding nodes $v_{S_i}$ in $\mathcal{T}$ must be colored red as $S_i$ would be the smallest witness set containing $C$.
}               
\end{figure}
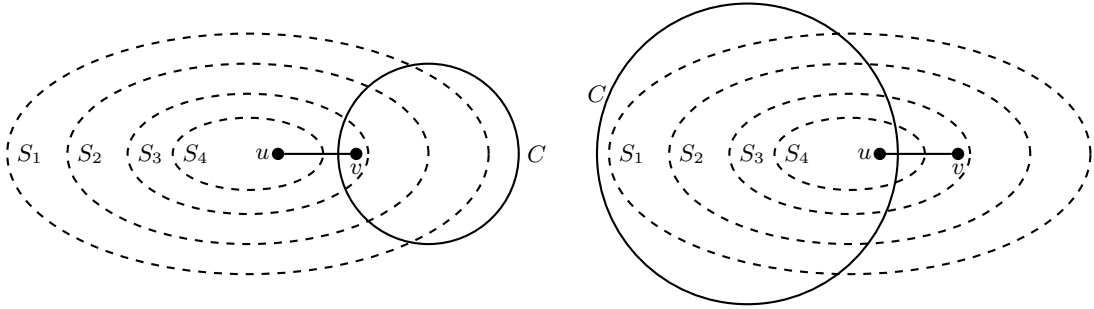
}   

\subsection{Proof of Theorem \ref{thm:crossdensity}}

We first decompose the tree $\mathcal{T}$ into vertex-disjoint paths as follows. Initially, our set of paths $\mathcal{P}$ is empty. While there exists a red node $v_{S_k}$ that is not included in some path in $\mathcal{P}$, consider the unique path $v_{S_1},v_{S_2},\ldots, v_{S_k}$ in $\mathcal{T}$ such that none of the nodes $v_{S_1},v_{S_2},\ldots, v_{S_{k-1}}$ is colored red, and the parent of $v_{S_1}$ either belongs to a path in $\mathcal{P}$ or is colored red or is $v_V$. Add this path to $\mathcal{P}$ and repeat. Note that at termination, every non-root node in $\mathcal{T}$ is part of a unique path in $\mathcal{P}$ since all leaves of $\mathcal{T}$ are colored red (by Lemma \ref{lem:colornodes}). Now consider a path $v_{S_1},v_{S_2},\ldots, v_{S_k}$ in $\mathcal{P}$. Here, the node $v_{S_k}$ is colored red. We will analyze the number of crossings between active sets and witness sets $S_1,\ldots, S_k$ and compare it with the number of tokens on this path.

\begin{claims}\label{clm:crossingpairs}
    Let $v_{S_1},v_{S_2},\ldots, v_{S_k}$ be a path in $\mathcal{P}$. Let $t$ be the number of tokens on this path. Then, the number of crossing pairs $(C,S_i)$ where $C\in\mathcal{C}$ and $1\leq i\leq k$ is at least $t-3$.
\end{claims}

\begin{proof}
    \textit{Case 1:} Suppose each node $v_{S_1},\ldots,v_{S_{k-1}}$ has one token. Let $e_{k-1}$ be the edge such that $\delta_F(S_{k-1}) = \{e_{k-1}\}$. Edge $e_{k-1}$ is incident to some active set $C$, and this active set must cross ${S_1},\ldots,{S_{k-2}}$ by Claim \ref{clm:longchain}. Since $v_{S_k}$ has at most two tokens, the number of tokens on this path is at most $k+1$ and the number of crossing pairs is at least $k-2$.\\
    \textit{Case 2:} Let $i$ be the largest index such that $v_{S_i}$ has two tokens and $i\leq k-1$ so that $v_{S_i}$ is not colored red. Let $e_{i}$ be the edge such that $\delta_F(S_{i}) = \{e_i\}$. Then, there are two active sets $C_1, C_2$ incident to each endpoint of $e_i$ respectively and by Claim \ref{clm:longchain}, $C_1$ crosses all the sets ${S_1},\ldots,{S_{i}}$ and $C_2$ crosses all the sets ${S_1},\ldots,{S_{i-1}}$. \\
    \indent\textit{Case 2.1:} Suppose $v_{S_k}$ has two tokens, then by Claim \ref{clm:longchain}, there exists an active set $C$ that crosses all the sets ${S_1},\ldots,{S_{k-1}}$. Note that $C$ could be the same as $C_1$ or $C_2$. But now, the number of tokens on the path is at most $2i + k-i + 1 = k+i +1$ and the number of crossing pairs is at least $i-1 + k-1 = k+i - 2$.\\
    \indent\textit{Case 2.2:} Suppose $v_{S_k}$ has one token. Let $e_{k-1}$ be the edge such that $\delta_F(S_{k-1}) = \{e_{k-1}\}$. As earlier, by Claim \ref{clm:longchain}, there exists an active set that crosses ${S_1},\ldots,{S_{k-2}}$. But now, the number of tokens on the path is at most $2i + k-i = k+i$ and the number of crossing pairs is at least $i-1 + k-2 = k+i - 3$.\qed
\end{proof}

For any path in $\mathcal{P}$, Claim \ref{clm:crossingpairs} shows that the number of tokens on this path is at most 3 plus the number of crossing pairs $(C, S)$ such that $C \in \mathcal{C}$ and $v_S$ is on this path. Since each path contains exactly one red node, and paths in $\mathcal{P}$ are vertex disjoint, this implies that the total number of tokens is bounded by $3n_R + a$ where $a$ is the number of crossing pairs $(S,C)$ where $S\in\mathcal{L}$ and $C\in\mathcal{C}$. This is upper bounded by $3|\mathcal{C}| + \rho|\mathcal{C}|$ and so $\sum_{C\in\mathcal{C}}|\delta_F(C)| \leq (3+\rho)|\mathcal{C}|$. Hence, Theorem \ref{thm:WGMV} implies that we have proven Theorem \ref{thm:crossdensity}.

\section{Improvements for $\gamma$-Pliable Families}\label{sec:gammapliable}

In this section, we improve the results by BCGI \cite{bansal2024improved} who showed that the WGMV primal-dual algorithm is a 16-approximation algorithm for the $\mathcal{F}$-augmentation problem when $\mathcal{F}$ is a $\gamma$-pliable family. Nutov \cite{nutov2024improved} refined their analysis to show a 10-approximation algorithm. We go further and prove theorem \ref{thm:8appx}.

\subsection{Proof of Theorem \ref{thm:8appx}}

\begin{proof}

Note that the discussions in Section \ref{sec:crossdensity} are valid for any pliable family and so are applicable here. In particular, we can still employ the Token Distribution Phase I and Claim \ref{clm:longchain} is also valid. We prove a couple more claims that are true in the case of $\gamma$-pliable families.

\begin{claims}\label{clm1}
    Let $v_{S_1},v_{S_2}$ be a directed path in $\mathcal{T}$ where $v_{S_1}$ is colored black and $v_{S_2}$ is not colored red. Then, at the end of the first phase of token distribution, $v_{S_2}$ has at most one token.
\end{claims}

\begin{proof}
    Consider the edge $e_2$ such that $\delta_F(S_2) = \{e_2\}$. It has one end-point $u$ in $S_2$ and another end-point $v$ in $S_1\setminus S_2$. If $v_{S_2}$ had two tokens, there must be an active set $C$ that contains $u$ and not $v$. This active set $C$ cannot be contained in $S_2$ since $v_{S_2}$ is not colored red. Indeed, there would have to be a third witness set $S$ such that $C\subseteq S \subseteq S_2$, with $e_2\in\delta(S)$. This contradicts property (ii) of witness sets in Lemma \ref{lem:laminarwitness}. Furthermore, the active set $C$ cannot be contained in $S_1$ since $v_{S_1}$ is not colored red. Hence, $C$ crosses both $S_1$ and $S_2$. Then, by property $\gamma$, $S_1\setminus (S_2 \cup C)$ is either empty or violated. However it contains the vertex $v$ and so must be violated. But then, it must contain an active set and so $S_1$ must contain another witness set disjoint from $S_2$. This would imply that $v_{S_1}$ is colored green which is a contradiction.\qed
\end{proof}

\begin{figure}[H]
    \centering
    \begin{tikzpicture}
        \draw[thick,dashed] (0,0) ellipse (3 and 2);\node at (0.6,2.2) {$S_1$};
        \draw[thick] (-1.5,0) circle (2.5); \node at (-4.3,0.3) {$C$};
        \draw[thick,dashed] (0,0) ellipse (1.6 and 1.2); \node at (-0.2,1.4) {$S_2$};
        \node[circle, draw, fill = black, inner sep=1pt] at (0.5,0) (node u) {}; \node at (0.5,-0.2) {\small u};
        \node[circle, draw, fill = black, inner sep=1pt] at (2,0) (node v) {}; \node at (2,-0.2) {\small v};
        \draw[thick] (node u) -- (node v);
    \end{tikzpicture}
    \caption{Illustration of the proof of Claim 4. $S_1\setminus (S_2\cup C)$ is non-empty and so must be violated by property $\gamma$. But then, $S_1$ contains violated sets other than $S_2$ and so $v_{S_1}$ must be colored green.}
    \label{fig:claim4}
\end{figure}

\begin{claims}\label{clm2}
    Let $v_{S_1}, v_{S_2}, v_{S_3}$ be a directed path in $\mathcal{T}$ where $v_{S_1}$ and $v_{S_2}$ are colored black. Then, an active set cannot cross all three sets $S_1, S_2$ and $S_3$.
\end{claims}

\begin{proof}
    Let $C$ be an active set that crosses all three sets $S_1, S_2, S_3$. Then by property $\gamma$, $S_2\setminus (S_3 \cup C)$ is either violated or empty. It cannot be violated as otherwise it would contain an active set and so $S_2$ would contain another witness set disjoint from $S_3$ forcing $v_{S_2}$ to be colored green. Similarly $S_1\setminus (S_2 \cup C)$ is empty. Hence, $S_2\setminus S_3 \subseteq C$ and $S_1\setminus S_2\subseteq C$. Consider the edge $e_2$ such that $\delta_F(S_2) = \{e_2\}$. It must have one end-point in $S_2\setminus S_3$ and the other end-point in $S_1\setminus S_2$. However both these end-points now lie inside the active set $C$. Since edge $e_2 \in\delta_F(C')$ for some active set $C'$, we have $C\cap C'$ is non-empty. This contradicts Lemma \ref{lem:colornodes}. \qed
\end{proof}

\begin{claims}\label{clm3}
    Let $v_{S_1},\ldots, v_{S_k}$ be a directed path in $\mathcal{T}$ with no red nodes. Then, there exists an active set that crosses all the sets $S_1,\ldots, S_{k-1}$.
\end{claims}

\begin{proof}
    This follows immediately from Claim \ref{clm:longchain}.
\end{proof}

Now define a \textbf{maximal chain} (or simply chain) in $\mathcal{T}$ to be a maximal directed path $v_{S_1},\ldots, v_{S_k}$ in $\mathcal{T}$ where $v_{S_k}$ is the only colored node. We call this a red (green) chain if $v_{S_k}$ is colored red (green). Observe that every colored node in $\mathcal{T}$ belongs to a unique chain and these chains do not intersect (since black nodes have outdegree one).

\begin{lemma}\label{lem:redchain}
    A red chain has a total of at most 5 tokens after the first phase of token distribution.
\end{lemma}

\begin{proof}
    We consider cases based on the length $k$ of the red chain $v_{S_1},\ldots, v_{S_k}$. Recall that each node has at most two tokens at the end of the first phase of token distribution.\\
    \textit{Case 1:} ($k\leq 2$). The chain has a total of at most 4 tokens.\\
    \textit{Case 2:} ($k = 3$). By Claim \ref{clm1}, node $v_{S_2}$ has at most one token and so the chain has at most 5 tokens in total.\\
    \textit{Case 3:} ($k=4$). By Claim \ref{clm1}, nodes $v_{S_2}$ and $v_{S_3}$ have at most one token each. We will show that $v_{S_4}$ also has at most one token, completing this case. Let $e_4 = (u,v)$ be the edge such that set $\delta_F(S_4) = \{e_4\}$, with $u\in S_4$ and $v\in S_{3}\setminus S_4$. Suppose $v_{S_4}$ has two tokens, then there exists an active set $C$ that contains $v$ and not $u$. Hence $C$ is not contained in $S_4$. Furthermore, $C$ is not contained in $v_{S_1}, v_{S_2}, v_{S_3}$ since they are not colored red. However, $C$ intersects all these sets since it contains $v$. This implies that the active set $C$ crosses all three sets $v_{S_1}, v_{S_2}, v_{S_3}$ contradicting Claim \ref{clm2}. Hence, $v_{S_4}$ has at most one token.\\
    \textit{Case 4:} ($k\geq 5$). This case cannot happen. Indeed by Claim \ref{clm3}, there exists an active set that crosses the sets $v_{S_1}, v_{S_2}, v_{S_3}$. But this contradicts Claim \ref{clm2}. \qed
\end{proof}

\begin{lemma}\label{lem:greenchain}
    Let $v_{S_1}, v_{S_2}, \ldots, v_{S_k}$ be a green chain. Then,
    \begin{enumerate}
        \item[(i)] $k \leq 3$.
        \item[(ii)] If $k = 3$, then the chain has at most $4$ tokens and the children of $v_{S_3}$ are all colored red.
        \item[(iii)] If $k \leq 2$, the the chain has at most $3$ tokens.
    \end{enumerate}
\end{lemma}

\begin{proof}
    \begin{enumerate}
        \item[(i)] If $k>3$, then by Claim \ref{clm3}, there exists an active set that crosses the sets $v_{S_1}, v_{S_2}, v_{S_3}$. But this contradicts Claim \ref{clm2}.
        \item[(ii)] If $k=3$, nodes $v_{S_2}$ and $v_{S_3}$ have at most one token each by Claim \ref{clm1} and so the chain has at most 4 token in total. Suppose $v_{S_3}$ had a child that is not colored red, then by Claim \ref{clm3}, there exists an active set that crosses the sets $v_{S_1}, v_{S_2}, v_{S_3}$. But this contradicts Claim \ref{clm2}.
        \item[(iii)] If $k=2$, then by Claim \ref{clm1}, node $v_{S_2}$ has at most one token.\qed
    \end{enumerate}
\end{proof}

We now perform a second phase of token distribution to leverage the lemmas above. \\

\noindent\fbox{\parbox{\textwidth}{\textit{Token Distribution Phase II}\\ Perform a root-to-leaf scan of $\mathcal{T}$. Upon encountering a black node, move all its tokens to its unique child. Upon encountering a green node whose chain has length three, evenly distribute all its tokens amongst its children.}}

~\\
By Lemma \ref{lem:greenchain}, green nodes whose chain has length three has children that are all red and these green chains have at most 4 tokens in total at the end of the first phase of token distribution. Hence, after the second phase of token distribution, these red children (there must be at least two) have at most 4 tokens each (2 initially and 2 more from its parent green node). All other red nodes have at most $5$ tokens each by Lemma \ref{lem:redchain} and all other green nodes have at most $3$ tokens each by Lemma \ref{lem:greenchain}. Thus, the total number of tokens is bounded by $5n_R + 3n_G$ which is at most $8|\mathcal{C}|$. Hence $\sum_{C\in\mathcal{C}}\delta_F(C) \leq 8|\mathcal{C}|$. Hence, Theorem \ref{thm:WGMV} implies that we have proven Theorem \ref{thm:8appx}.
\end{proof}

\section{Small Cut Augmentation}\label{sec:SmallCuts}

In this section, we prove Theorem \ref{thm:SmallCuts}. We will build on the analysis of the previous Section \ref{sec:gammapliable} and utilize the sparse crossing property defined in Definition \ref{def:sparsecrossing} to improve on the approximation factor. We begin by showing that the family of small cuts satisfies the sparse crossing property.

\begin{lemma}\label{lem:smallcutssparsecrossing}
    The family of small cuts satisfies the sparse crossing property.
\end{lemma}

\begin{proof}
    Let $\mathcal{F}$ be a small cut family. Let $D\subseteq E$ and $S \in \mathcal{F}_D$ and suppose that $S$ crosses two minimal sets of $\mathcal{F}_D$ say $C_1$ and $C_2$. Define $u(T_1,T_2)$ for sets $T_1,T_2\subseteq V$ to be $\sum_{e\in E_0(T_1,T_2)} u_e$ where $E_0(T_1,T_2)$ is the set of edges in $E_0$ with exactly one end point in $T_1$ and exactly the other end point in $T_2$. Also define $u(T) = u(T,\overline{T})$. The following equation is easy to verify:
    \[
    u(C_1\setminus S) + u(S\cap C_1) = u(C_1) + 2u(S\cap C_1,C_1\setminus S)
    \]
    Since $C_1$ is minimal in $\mathcal{F}_D$, we can conclude that $S\cap C_1$ and $C_1\setminus S$ are not in $\mathcal{F_D}$. Hence, $u(S\cap C_1)$ and $u(C_1\setminus S)$ are both at least the threshold $k$. Furthermore $u(C_1) < k$ since $C_1\in\mathcal{F}$. Hence, $u(S\cap C_1,C_1\setminus S) \geq k/2$. Similarly $u(S\cap C_2, C_2\setminus S) \geq k/2$. Now since $C_1$ and $C_2$ are disjoint (By Lemma \ref{lem:colornodes}), we can conclude that $u(S) \geq k$. But this contradicts the fact that $S\in\mathcal{F}$. Hence $S$ cannot cross two minimal sets $C_1$ and $C_2$.\qed
\end{proof}

We now present a proof of Theorem \ref{thm:6-appx} showing that the WGMV algorithm is a 6-approximation algorithm for the $\mathcal{F}$-augmentation problem when $\mathcal{F}$ is a $\gamma$-pliable family satisfying the sparse crossing property.

\subsection{Proof of Theorem \ref{thm:6-appx}}

\begin{proof}
    Note that claims \ref{clm1},\ref{clm2},\ref{clm3} are all valid here as well since they are valid for all $\gamma$-pliable families. We will prove a couple more claims and lemmas and then apply a final third round of token distribution after which red nodes will have at most 6 tokens each and every other node will not have tokens.

\begin{claims}\label{clm4}
    Let $v_{S_1}, v_{S_2}$ be a directed path in $\mathcal{T}$ such that $v_{S_2}$ has two tokens at the end of the first phase of distribution. Then, at least one of $v_{S_1}$ and $v_{S_2}$ is colored red.
\end{claims}

\begin{proof}
    Suppose neither of $v_{S_1}$ nor $v_{S_2}$ is colored red. Let $e_2 = (u,v)$ be the edge such that $\delta_F(S_2) = \{e_2\}$, with $u\in S_2$ and $v\in S_1\setminus S_2$. Since $v_{S_2}$ has two tokens, there exists active sets $C_u$ and $C_v$ incident to edge $e_2$ that contain $u$ and $v$ respectively. These active sets are not contained in $S_1$ nor $S_2$ (since they are not colored red). However, they both intersect set $S_1$. But then, the violated set $S_1$ crosses two active sets $C_u$ and $C_v$ contradicting the sparse crossing property.\qed
\end{proof}

\begin{lemma}\label{lem:greenchainnew}
    Let $v_{S}$ be the child of a green node $v_{T}$ in $\mathcal{T}$. Then,
    \begin{enumerate}
        \item[(i)] If $v_{S}$ belongs to a red chain, then this red chain has at most 4 tokens after the second phase of token distribution.
        \item[(ii)] If $v_{S}$ belongs to a green chain, then this green chain has at most 2 tokens after the second phase of token distribution. 
    \end{enumerate}
\end{lemma}

\begin{proof}
    Let $v_{S_1}, v_{S_2},\ldots,v_{S_k}$ be the unique chain with $v_S = v_{S_1}$.
    \begin{enumerate}
        \item[(i)] Suppose $v_{S_k}$ is colored red. If $v_{S_k}$ received additional tokens in the second phase of redistribution from a green node, then $v_{S_k}$ has at most 4 tokens as shown in the proof of Theorem \ref{thm:8appx}. So assume that $v_{S_k}$ did not receive additional tokens in the second phase of redistribution from a green node. If $k\leq 2$, then $v_{S_k}$ trivially has at most 4 tokens. If $k\geq 3$, then Lemma \ref{lem:redchain} showed that $v_{S_k}$ has at most 5 tokens. However $v_{S_k}$ having 5 tokens required $v_{S_1}$ to have 2 tokens after the first phase of token distribution. This is ruled out by Claim \ref{clm4} applied to the path $v_T, v_{S_1}$ where none of these nodes is red. Hence, $v_{S_k}$ has at most 4 tokens.
        \item[(ii)] Suppose $v_{S_k}$ is colored green. If $k \geq 3$, then after the second phase of token distribution, $v_{S_k}$ has no tokens. If $k = 2$, then $v_{S_1}$ and $v_{S_2}$ have at most one token after the first phase (By Claim \ref{clm4} applied to the paths $v_T, v_{S_1}$ and $v_{S_1},v_{S_2}$ where none of these nodes is red). Hence, $v_{S_k}$ has at most 2 tokens. The case $k=1$ is trivial.\qed
    \end{enumerate} \qed
\end{proof}

Now, define a \textit{maximal green subtree} of $\mathcal{T}$ to be a maximal subtree rooted at a node $v$ that is colored green such that all leaves of this subtree are colored red and no non-leaf node is colored red. Observe that every green node $v$ of $\mathcal{T}$ belongs to a unique maximal green subtree $\mathcal{T}_v$ of $\mathcal{T}$ and every red node of $\mathcal{T}$ belongs to at most one maximal green subtree of $\mathcal{T}$. Perform now, the following third phase of token distribution,\\

\noindent\fbox{\parbox{\textwidth}{\textit{Token Distribution Phase III}\\ For every maximal green subtree of $\mathcal{T}$, evenly distribute all tokens from its green nodes to its red nodes.}}

~\\Let's analyze the the subtree $\mathcal{T}_v$. The number of red nodes in $\mathcal{T}_v$ is at least the number of green nodes in $\mathcal{T}_v$ plus one. By Lemma \ref{lem:greenchainnew}, each of these red nodes has at most 4 tokens at the end of the second phase and every non-root green node has at most 2 tokens. The green root node has at most 3 tokens after the second phase. Hence, this third phase of token distribution ensures that no red node gains more than 2 additional tokens. Since every red node of $\mathcal{T}$ belongs to at most one maximal green subtree, red nodes have at most 6 tokens after this third phase and green nodes have no tokens. Now Theorem \ref{thm:WGMV} implies that we have proven Theorem \ref{thm:6-appx}. \qed

\end{proof}

Thus finally, Theorem \ref{thm:6-appx} along with Lemma \ref{lem:smallcutssparsecrossing} proves Theorem \ref{thm:SmallCuts} and provides the current best approximation factor for the Small Cut Augmentation problem. 

~\\~\\~\\~
In this section we tackle the question of how one might prove that the crossing density of set family is bounded. Note that while Definition \ref{def:CrossingDensity} of the crossing density considered all laminar subfamilies $\mathcal{F}_D^{lam}$, the proof of Theorem $\ref{thm:crossdensity}$ required this only for the laminar family formed by the witness sets. Hence, we shall bound the crossing density with respect to just the laminar family of witness sets i.e. we will bound the ratio
\begin{equation}\label{eq:crossingdensity} \frac{\text{number of crossing pairs $(S,C)$ where $S\in\mathcal{F}_D^{wit}$ and $C\in\mathcal{C}_D$ }} {|\mathcal{C}_D|}
\end{equation}
where $\mathcal{F}_D^{wit} = \mathcal{L}$ is the laminar witness family given by Lemma \ref{lem:laminarwitness}. To do so, we introduce a new local property of set families that we call the \textbf{sparse crossing} property.

\begin{definition}[Sparse Crossing Property]
    A set family $\mathcal{F}$ has the sparse crossing property if for any subset of the edges $D\subseteq E$, a set $S\in\mathcal{F}_D$ crosses at most one set in $\mathcal{C}_D$.
\end{definition}

As we will shortly see, the family of small cuts satisfies the sparse crossing property. As one might imagine, such a property should be useful in bounding the crossing density of a set family. We show precisely this via the following theorem.

\begin{theorem}\label{thm:sparsecrossing}
    Let $\mathcal{F}$ be a $\gamma$-pliable set family that also satisfies the sparse crossing property. Then the crossing density \eqref{eq:crossingdensity} is bounded by 2.
\end{theorem}

\begin{proof}
    First, consider an active set $C\in\mathcal{C}_D$ that crosses every set in a chain $S_1 \subset S_2 \subset \cdots \subset S_k$ where $S_i$ are sets in $\mathcal{L}$. Then, we claim that for $i=3,5,7,\ldots$, there exists an active set in $S_{i}\setminus (S_{i-2} \cup C)$. Indeed by property $\gamma$, either $S_i \setminus (C\cup S_{i-1})$ contains an active set, or $S_{i-1}\setminus (C\cup S_{i-2})$ contains an active set, or both these sets are empty. However in the latter case where both the sets are empty, we see that $S_{i}\setminus S_{i-1}$ and $S_{i-1}\setminus S_{i-2}$ are both contained in $C$. However, the edge $e_{i-1}$ such that $ \{e_{i-1}\}= \delta_F(S_{i-1})$ has one end-point in $S_{i}\setminus S_{i-1}$ and the other in $S_{i-1}\setminus S_{i-2}$. This would imply that edge $e_{i-1}$ is contained inside $C$ which is a contradiction since $e_{i-1}$ is then incident to an active set that intersects $C$ (contradicting Lemma \ref{lem:colornodes}).

    We now assign charges to some active sets and will compare the number of charges distributed to the number of crossing pairs. First, we define a partial order on active sets using the rooted tree $\mathcal{T}$. For each active set $C$, we assign to it the smallest set in the laminar family $\mathcal{L}$ that contains $C$. Suppose $S_1$ is assigned to $C_1$ and $S_2$ is assigned to $C_2$. We say that $C_1 \succ C_2$ if and only if the node $v_{S_1}$ is an ancestor of the node $v_{S_2}$ in the tree $\mathcal{T}$. Now consider an active set $C$ and a maximal chain $S_1 \subset S_2 \subset \cdots \subset S_k$ that $C$ crosses. Note that $S_1\setminus C$ is in $\mathcal{F}_D$ by the pliable property and the minimality of $C$ and hence contains an active set. While breaking ties arbitrarily, we assign one charge to a largest (as given by $\succ$) active set contained in $S_1\setminus C$ and a largest active set contained in $S_{i}\setminus (S_{i-2}\cup C)$ for $i=3,5,7,\ldots$.

    We claim that every active set has at most one charge currently. Suppose that there exists an active set $C^*$ that contains two charges. Then $C^*$ receives a charge due to the crossing between $C$ and a chain $S_1 \subset S_2 \subset \cdots \subset S_k$ and also due to the crossing between $C'$ and a chain $S'_1 \subset S'_2 \subset \cdots \subset S'_{k'}$. This implies that $C^*$ is contained in $S_i\setminus C$ for some $i$ and also $S'_j\setminus C'$ for some $j$. WLOG assume that $S'_j \subseteq S_i$. Due to the sparse crossing property, $S'_j$ cannot cross $C'$ and $C$ and so is strictly contained in $S_1$. This implies that $C^*$ is also contained in $S_1$. Since $C^*$ receives a charge from the crossing between $C$ and the chain $S_1 \subset S_2 \subset \cdots \subset S_k$, we can conclude that $C^*$ is a largest active set contained in $S_1\setminus C$. But now consider the active set $C'$ that crosses $S'_j$. This cannot cross $S_1$ (nor $C$) by the sparse crossing property and so is contained in $S_1\setminus C$ but not in $S'_j$. But then $C' \succ C^*$ contradicting the observation that $C^*$ is a largest active set contained in $S_1\setminus C$ (See Figure \ref{fig:claim3}).
\begin{figure}[H]
    \centering
    \begin{tikzpicture}
        \draw[thick,dashed] (0,0) ellipse (3 and 2);\node at (-0.2,2.2) {$S_1$};
        \draw[thick] (-3,0.3) circle (1); \node at (-3.3,0.3) {$C$};
        \draw[thick,dashed] (0,0) ellipse (1.6 and 1.2); \node at (-0.2,1.4) {$S_j'$};
        \draw[thick] (0,0) circle (0.6);\node at (0,0) {$C^*$};
        \draw[thick] (1.5,0.6) circle (0.6); \node at (1.6,0.7) {$C'$};
    \end{tikzpicture}
    \caption{Illustration of the proof that active sets have at most one charge. $C' \succ C^*$ and so $C^*$ could not have received a charge from the crossing between $S_1$ and $C$.}
    \label{fig:claim3}
\end{figure}
Hence, the number of active sets is at least equal to the number of charges distributed. We now argue that the number of charges distributed is at least half of the number of crossing pairs $(S,C)$ where $S$ is a witness set in $\mathcal{L}$ and $C$ is an active set in $\mathcal{C}_D$. Indeed for any maximal chain $S_1 \subset S_2 \subset \cdots \subset S_k$ that an active set $C$ crosses, we have distributed at least $k/2$ charges. Furthermore, by the sparse crossing property, no witness set in this chain crosses any other active set. Hence, the number of charges distributed is at least half the number of crossing pairs. This implies that the crossing density \eqref{eq:crossingdensity} is bounded by 2.\qed
\end{proof}

To prove Theorem \ref{thm:SmallCuts} now, all we have to show is that the family of small cuts is a $\gamma$-pliable family that satisfies the sparse crossing property. For then, Theorem \ref{thm:sparsecrossing} and Theorem \ref{thm:crossdensity} will imply that the WGMV primal-dual algorithm is a 5-approximation algorithm for the small cut augmentation problem.

\subsubsection{Proof of Theorem \ref{thm:SmallCuts}}
    It was shown in \cite{bansal2024improved} that the family of small cuts $\mathcal{F}$ is $\gamma$-pliable. Hence we only need to show the sparse crossing property. Let $D\subseteq E$ and $S \in \mathcal{F}_D$ and suppose that $S$ crosses two minimal sets of $\mathcal{F}_D$ say $C_1$ and $C_2$. Define $u(T_1,T_2)$ for sets $T_1,T_2\subseteq V$ to be $\sum_{e\in E_0(T_1,T_2)} u_e$ where $E_0(T_1,T_2)$ is the set of edges in $E_0$ with exactly one end point in $T_1$ and exactly the other end point in $T_2$. Also define $u(T) = u(T,\overline{T})$. The following equation is easy to verify:
    \[
    u(C_1\setminus S) + u(S\cap C_1) = u(C_1) + 2u(S\cap C_1,C_1\setminus S)
    \]
    Since $C_1$ is minimal in $\mathcal{F}_D$, we can conclude that $S\cap C_1$ and $C_1\setminus S$ are not in $\mathcal{F_D}$. Hence, $u(S\cap C_1)$ and $u(C_1\setminus S)$ are both at least the threshold $k$. Furthermore $u(C_1) < k$ since $C_1\in\mathcal{F}$. Hence, $u(S\cap C_1,C_1\setminus S) \geq k/2$. Similarly $u(S\cap C_2, C_2\setminus S) \geq k/2$. Now since $C_1$ and $C_2$ are disjoint (By Lemma \ref{lem:colornodes}), we can conclude that $u(S) \geq k$. But this contradicts the fact that $S\in\mathcal{F}$. Hence $S$ cannot cross two minimal sets $C_1$ and $C_2$.\qed

\section{Flexible Graph Connectivity}\label{sec:FGC}

In this section, we prove Theorem \ref{thm:FGC} exhibiting the first constant factor approximation algorithm for $(p,3)$-FGC. The cases when $p$ is even and odd are handled separately. We begin by introducing some notation that will be useful in the discussions to follow.

\subsection{Notation and Terminology}

We will refer to the set of unsafe edges as $\mathbf{\mathcal{U}}$. A cut of a graph is simply $\delta(A)$ for some subset of the vertices $A\subseteq V$, and when the context is clear we might refer to the cut as $A$ instead of $\delta(A)$. For $A,B\subseteq V$ with $A \cap B = \emptyset$, we define $\mathbf{E(A,B)}$ to be the set of edges with one end point in $A$ and the other in $B$.

Let $\mathcal{F}$ be a family of cuts of a graph $G$. We define an equivalence relation between vertices of the graph $G$ as follows: $\mathbf{u\equiv v}$ if and only if for all $C\in\mathcal{F}$, either $u,v \in C$ or $u,v\not\in C$. The residual graph of $G$ with respect to $\mathcal{F}$ is obtained by contracting each equivalence class of vertices into a single vertex. We will call this graph $\mathbf{G_{\mathcal{F}} = (V_{\mathcal{F}},E_{\mathcal{F}})}$. We will call a vertex (edge) in $G_{\mathcal{F}}$ a \textbf{residual vertex (edge)}. Note that any residual vertex (edge) is a subset of vertices (edges) in the original graph $G$. Hence the cardinality of a residual edge makes sense and so does the cardinality of the unsafe edges in a residual edge. We will call a residual edge \textbf{thin} if it contains at most one unsafe edge.

For two sets $A,B\subseteq V$, we refer to the cuts $\delta(A\setminus B), \delta(A\cap B),\delta(B\setminus A), \delta(A\cup B)$ as \textbf{corner cuts}. In the graph $G_{\{A,B\}}$, we refer to the residual edges $(A\cap B,\overline{A\cup B})$ and $(A\setminus B,B\setminus A)$ as \textbf{diagonal edges}. We refer to the residual edges $(A\cap B,A\setminus B)$ and $(A\cap B,B\setminus A)$ as \textbf{internal edges} and the residual edges $(\overline{A\cup B},A\setminus B)$ and $(\overline{A\cup B},B\setminus A)$ as \textbf{external edges}. We say that a set $A\subseteq V$ \textbf{amenably crosses} a set $B\subseteq V$ if (i) $A$ crosses $B$ and (ii) either the residual internal edge $(A\cap B,B\setminus A)$ or the residual external edge $(A\setminus B, \overline{A\cup B})$ does not contain any unsafe edges.

We say that a graph $G = (V,E)$ is $\mathbf{(p,q)}$\textbf{-flex-connected} if for any subset of the unsafe edges $F\subseteq \mathcal{U}$ with $|F|\leq q$, the graph $G = (V,E\setminus F)$ is $p$-edge connected. For a graph $G$, we call a cut $\mathbf{(p,q)}$\textbf{-violated} if it contains exactly $p+q-1$ edges, at least $q$ of which are unsafe.

\subsection{Approximation Algorithms}

We employ the augmentation framework for flexible graph connectivity utilized by \cite{boyd2024approximation,bansal2024improved,chekuri2023,nutov2023improved}. We start with a solution $H=(V,F)$ that is $(p,2)$-flex-connected using the 6-approximation algorithm in \cite{bansal2024improved} when $p$ is even and the $7+\epsilon$-approximation algorithm in \cite{nutov2023improved} when $p$ is odd. At this point, each cut either has $p$ safe edges or $p+2$ edges in total. Hence, the only cuts which require more edges to make $H$ $(p,3)$-flex-connected are cuts with fewer than $p$ safe edges and exactly $p+2$ edges in total. These are precisely the $(p,3)$-violated cuts and we will refer to these as simply violated cuts henceforth. Furthermore, adding just a single edge (safe or unsafe) to these violated cuts will ensure that they have $p+3$ edges in total and so $H$ would become $(p,3)$-flex-connected. Thus, we only need to solve the $\mathcal{F}$-augmentation problem where $\mathcal{F}$ is the family of $(p,3)$-violated cuts of a $(p,2)$-flex-connected graph $H$. We prove the following theorems.

\begin{theorem}\label{thm:p_even}
    Let $p\geq 2$ be even and let $G = (V,E)$ be a $(p,2)$-flex-connected graph. Then the family of violated cuts forms a $\gamma$-pliable family with the sparse crossing property.
\end{theorem}

\begin{theorem}\label{thm:p_odd}
    Let $p\geq 3$ be odd let $G = (V,E)$ be a $(p,2)$-flex-connected graph. Let $\mathcal{F}$ be the family of violated cuts and let $\mathcal{F}_1 \subset \mathcal{F}$ be the family violated cuts that either $(i)$ do not cross any other violated set or $(ii)$ cross at least one other violated set amenably. Then, $\mathcal{F}_1$ and $\mathcal{F}\setminus\mathcal{F}_1$ are both uncrossable families.
\end{theorem}

Theorem \ref{thm:p_even} allows us to augment the family of violated cuts incurring an approximation factor of $6$ via Theorem \ref{thm:6-appx}. Thus, we obtain a $12$-approximation algorithm for $(p,3)$-FGC when $p$ is even. Theorem \ref{thm:p_odd} allows us to augment each of the families $\mathcal{F}_1$ and $\mathcal{F}\setminus\mathcal{F}_1$ incurring an approximation factor of 2 each. Thus, we obtain an $11+\epsilon$-approximation algorithm for $(p,3)$-FGC when $p$ is odd.

\subsubsection{Proof of Theorem \ref{thm:p_even}}~

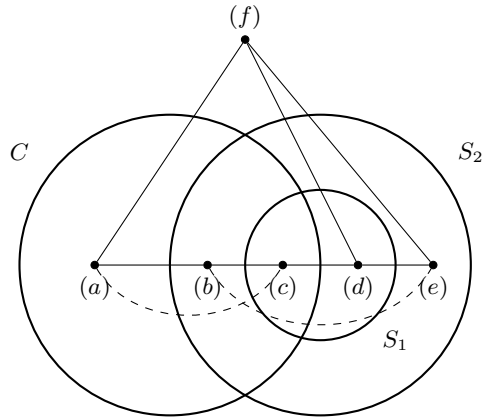
\begin{figure}
        \centering
        \begin{tikzpicture}
            \draw[thick] (0,0) circle (2cm); \node at (-2,1.5) {$C$};
        \draw[thick] (2,0) circle (2cm); \node at (4,1.5) {$S_2$};
        \draw[thick] (2,0) circle (1cm); \node at (3,-1) {$S_1$};
        \node[circle, draw, fill = black, inner sep=1pt] at (-1,0) (node a) {}; \node at (-1,-0.3) {\small $(a)$};
        \node[circle, draw, fill = black, inner sep=1pt] at (0.5,0) (node b) {}; \node at (0.5,-0.3) {\small $(b)$};
        \node[circle, draw, fill = black, inner sep=1pt] at (1.5,0) (node c) {}; \node at (1.5,-0.3) {\small $(c)$};
        \node[circle, draw, fill = black, inner sep=1pt] at (2.5,0) (node d) {}; \node at (2.5,-0.3) {\small $(d)$};
        \node[circle, draw, fill = black, inner sep=1pt] at (3.5,0) (node e) {}; \node at (3.5,-0.3) {\small $(e)$};
        \node[circle, draw, fill = black, inner sep=1pt] at (1,3) (node f) {}; \node at (1,3.3) {\small $(f)$};
        \draw (node a) -- (node b) -- (node c) -- (node d) -- (node e) -- (node f) -- (node a);
        \draw[bend right = 60,dashed] (node a) to (node c);
        \draw[bend right = 60,dashed] (node b) to (node e);
        \draw (node d) -- (node f);  
        \end{tikzpicture}
        \caption{$C$ is minimal and crosses $S_1\subseteq S_2$. Then the residual vertex $(e)$ must be violated or empty.}
        \label{fig:propertygamma}
    \end{figure}

    We begin by analyzing the residual graph of two crossing violated sets $A,B$. Recall that $A$ and $B$ are both $p+2$-cuts with at least three unsafe edges each. The following equations about multisets are well known and can be easily verified:
    \begin{align*}
        \delta(A) + \delta(B) &= \delta(A\cup B) + \delta(A\cap B) + 2E(A\setminus B, B\setminus A) \tag{a}\label{eq:a}\\
        \delta(A) + \delta(B) &= \delta(A\setminus B) + \delta(B\setminus A) + 2E(\overline{A\cup B}, A\cap B)\tag{b}\label{eq:b}\\
        \delta(A) + 2E(A\setminus B, A\cap B) &= \delta(A\setminus B) + \delta(A\cap B)\tag{c}\label{eq:c}\\
        \delta(B) + 2E(B\setminus A, A\cap B) &= \delta(B\setminus A) + \delta(A\cap B) \tag{d}\label{eq:d}\\
        \delta(A) + 2E(B\setminus A, \overline{A\cup B}) &= \delta(B\setminus A) + \delta(A\cup B)\tag{e}\label{eq:e}\\
        \delta(B) + 2E(A\setminus B, \overline{A\cup B}) &= \delta(A\setminus B) + \delta(A\cup B)\tag{f}\label{eq:f}
    \end{align*}
    Since $A$ and $B$ cross, the four residual vertices are all non-empty. Since the graph $G$ is $(p,2)$-flex-connected, the corresponding corner cuts $\delta(A\setminus B),\delta(A\cap B),\delta(B\setminus A),\delta(A\cup B)$ have cardinality at least $p$. Equation \eqref{eq:a} tells us that the pair of $\{|\delta(A\cup B)|,|\delta(A\cap B)|\}$ have the same parity and lie in the interval $\{p,\ldots,p+4\}$ and sum to at most $2p+4$. Equation \eqref{eq:b} tells us the same about the pair $\{|\delta(A\setminus B)|,|\delta(B\setminus A)|\}$. Furthermore equation \eqref{eq:c} tells us that these pairs have same parities. Hence, the sizes of all corner cuts lie in either $\{p,p+2,p+4\}$ or in $\{p+1,p+3\}$. Now suppose two corner cuts had size $p+1$ or two corner cuts had size $p$, then these two cuts together can have at most two unsafe edges ($G$ is $(p,2)$-flex-connected). This would contradict at least one of the equations \eqref{eq:a},\eqref{eq:b},\eqref{eq:c},\eqref{eq:d},\eqref{eq:e},\eqref{eq:f} since the number of unsafe edges in the LHS and RHS of these equations will not match. This is simple to see for equations \eqref{eq:c}-\eqref{eq:f}. For \eqref{eq:a} (or \eqref{eq:b}), we will obtain that the RHS has at most 4 unsafe edges (as $|E(A\setminus B, B\setminus A)| = 1$) in this case) while the LHS has at least 6 unsafe edges. With this observation, we can conclude that the four corner cuts have sizes in $\{p,p+2,p+4\}$ and there is at most one $p$-cut. We show that the family of violated cuts forms a pliable family by considering the following list of exhaustive cases:
    \begin{description}
        \item[Case 1:] $\delta(A\setminus B)$ is a $p$-cut. Then, $\delta(A\cap B)$ and $\delta(A\cup B)$ must be $p+2$-cuts (since otherwise one of them will be a $p$-cut). Notice that $\delta(A\setminus B)$ cannot contain any unsafe edges since $G$ is $(p,2)$-flex connected. Hence the unsafe edges of $\delta(A)$ are all in $\delta(A\cap B)$ and the unsafe edges of $\delta(B)$ are all in $\delta(A\cup B)$. This implies that $A\cup B$ and $A\cap B$ are also violated.\\
        \item[Case 2:] $\delta(A\cup B)$ is a $p$-cut. Similar to case 1, $\delta(A\setminus B)$ and $\delta(B\setminus A)$ must be $p+2$-cuts and every unsafe edge of $\delta(A)$ must be in $\delta(B\setminus A)$ and every unsafe edge of $\delta(B)$ must be in $\delta(A\setminus B)$. This implies that $A\setminus B$ and $B\setminus A$ are also violated.\\
        \item[Case 3:] $\delta(A\cap B)$ is a $p$-cut. Similar to case 2, $\delta(A\setminus B)$ and $\delta(B\setminus A)$ must be $p+2$-cuts and every unsafe edge of $\delta(A)$ must be in $\delta(A\setminus B)$ and every unsafe edge of $\delta(B)$ must be in $\delta(B\setminus A)$. This implies that $A\setminus B$ and $B\setminus A$ are also violated.\\
        \item[Case 4:] All four corner cuts are $(p+2)$-cuts. In this case, there are no diagonal edges in the residual graph and every edge in the residual graph has size $p/2 + 1$. By equation \eqref{eq:a}, at least one of the cuts $\delta(A\cup B)$ or $\delta(A\cap B)$ must have at least three unsafe edges and by \eqref{eq:b}, at least one of the cuts $\delta(A\setminus B)$ or $\delta(B\setminus A)$ must have at least three unsafe edges. This implies that at least two of the four corner cuts are violated.
    \end{description}
    We have exhausted all the cases and shown that the family of violated cuts forms a pliable family. We now argue that the family of violated cuts satisfies property $\gamma$. Let $D\subseteq E$ be a set of edges and let $C, S_1, S_2 \in \mathcal{F}_D$ such that $C$ is inclusion-wise minimal, $S_1,S_2$ cross $C$ and $S_1\subseteq S_2$. Assume for the sake of contradiction that $S_2\setminus C\cup S_1$ is neither empty, nor violated. Since $C$ and $S_1$ are violated sets, at least two of the four corner cuts $\delta(C\cap S_1), \delta(C\cup S_1),\delta(C\setminus S_1)$ and $\delta(S_1\setminus C)$ must also be violated (since $\mathcal{F}$ is pliable). If any of these corner cuts intersects $D$, then either $C$ or $S_1$ also intersects $D$ (contradicting $C,S_1\in\mathcal{F}_D$). Furthermore, since $C$ is a minimal set in $\mathcal{F}_D$, it must be the case that $C\cup S_1$ and $S_1\setminus C$ are in $\mathcal{F}_D$. As we observed from the case analysis above, this only happens in the fourth case when there are no diagonal edges in the residual graph of $\{C,S_1\}$ and every residual edge has size $p/2+1$. The same holds for the crossing between $C$ and $S_2$. The residual graph of $\{C,S_1,S_2\}$ is shown in figure \ref{fig:propertygamma}. All other edges are diagonal edges for some pair of crossing sets among $C,S_1,S_2$.
    The edge $(cd)$ is the internal edge of $\delta(C)$ with respect to $\delta(S_1)$ and the edges $(cd)$ and $(be)$ are the internal edges of $\delta(C)$ with respect to $\delta(S_2)$. Since these internal edges have the same sizes ($p/2+1$), the residual edge $(be)$ is empty. The edges $(fd)$ and $(de)$ are the external edges of $\delta(S_1)$ in the crossing with $C$ and so $|(fd)|+|(de)| = p/2+1$. Similarly, the edges $(fd)$ and $(fe)$ are the external edges of $\delta(S_2)$ in the crossing with $C$ and so $|(fd)|+|(fe)| = p/2+1$. This implies that $|(fe)| = |(de)|$. Since the residual vertex $(e)$ is non-empty and $G$ is $(p,2)$-flex-connected, we know $|(fe)|+|(de)| \geq p$. Hence $|(fd)|\leq 1$. Suppose $|(fd)| = 1$, then $|(fe)| = |(de)| = p/2$ and $\delta(e)$ is a $p$-cut containing no unsafe edges. But now, $C\cup S_1$ is violated implies that the residual edge $(af)$ contains at least 2 unsafe edges and $S_2$ is violated implies that the residual edges $(ab)$ and $(ac)$ contain at least two unsafe edges in total. But then the cut $\delta(a)$ will become violated since it is a $(p+2)$-cut with at least 3 unsafe edges. This contradicts the minimality of $C$. Hence $|(fd)| = 0$ and so $|(fe)| = |(de)| = p/2+1$ and $\delta(e)$ is a $(p+2)$-cut. Now suppose the number of unsafe edges in the edges $(fe)$ and $(de)$ are at most two in total. These are the external edges of the violated cuts $\delta(S_1)$ and $\delta(S_2)$ with respect to the crossing with $C$. Hence the internal edges must have the remainder of the unsafe edges. Suppose $|(fe)\cap\mathcal{U}| = 0$ or $|(de)\cap\mathcal{U}| = 0$, then either $|(af)\cap\mathcal{U}| \geq 3$ or $|((ac)\cup (bc) )\cap\mathcal{U}| \geq 3$. But then either $\delta(a)$ or $\delta(c)$ is violated contradicting the minimality of $C$. Thus, we can assume that $(fe)$ and $(de)$ contain exactly one unsafe edge each. But then $(cd)$ contains at least two unsafe edges as $S_1\setminus C$ is violated and $(ac)\cup (bc)$ contains at least two unsafe edges since $S_1$ is violated. But then $\delta(c)$ is violated contradicting the minimality of $C$. This implies that $(fe)$ and $(de)$ together contain at least three unsafe edges and so the residual vertex $(e) = S_2\setminus (S_1\cup C)$ is a violated set.
    
    Finally, we argue that the family of violated cuts satisfies the sparse crossing property. Let $D\subseteq E$ and $S \in \mathcal{F}_D$ and suppose that $S$ crosses two minimal sets of $\mathcal{F}_D$ say $C_1$ and $C_2$. The above discussions show that $|\delta(C_i\setminus S)| = |\delta(S\cap C_i)| = p+2$. Since these two cuts are not in $\mathcal{F}$ (by minimality of $C_i$), they have at most two unsafe edges each. Also, $\delta(C_i)$ has at least 3 unsafe edges and so $E(C_i\setminus S, S\cap C_i)$ has no unsafe edges by the cut equation \eqref{eq:c} above. By the same cut equation \eqref{eq:c}, $|E(C_i\setminus S, S\cap C_i)| = (p+2)/2$ and so due to cardinality constraints, $\delta(S) = E(C_1\setminus S, S\cap C_1) \cup E(C_2\setminus S, S\cap C_2)$. However, now $\delta(S)$ has no unsafe edges and this is a contradiction. This completes the proof. \qed

~\\We now move on to the case when $p$ is odd. This is more complicated since the family of violated cuts does not adhere to any recognizable structures that allows us to solve the $\mathcal{F}$-augmentation problem efficiently. Instead, we split this family into two uncrossable families by providing a careful and in-depth analysis of the family of violated cuts.

\subsubsection{Proof of Theorem \ref{thm:p_odd}}

    We begin by analyzing the residual graph of two crossing violated sets $A,B$. Recall that $A$ and $B$ are both $p+2$-cuts with at least three unsafe edges each. The equations \eqref{eq:a},\eqref{eq:b},\eqref{eq:c},\eqref{eq:d},\eqref{eq:e},\eqref{eq:f} are useful here.\\
    Since $A$ and $B$ cross, the four residual vertices are all non-empty. Since the graph $G$ is $(p,2)$-flex-connected, the corresponding corner cuts $\delta(A\setminus B),\delta(A\cap B),\delta(B\setminus A),\delta(A\cup B)$ have cardinality at least $p$. Equation \eqref{eq:a} tells us that the pair of $\{|\delta(A\cup B)|,|\delta(A\cap B)|\}$ have the same parity and lie in the interval $\{p,\ldots,p+4\}$ and sum to at most $2p+4$. Equation \eqref{eq:b} tells us the same about the pair $\{|\delta(A\setminus B)|,|\delta(B\setminus A)|\}$. Furthermore, since $p$ is odd, equation \eqref{eq:c} tells us that these pairs have different parities. Hence, one of these pairs is in $\{p,p+2,p+4\}$ and the other pair is in $\{p+1,p+3\}$. Therefore, at least one of the corner cuts has size $p+1$. Now suppose another corner cut had size at most $p+1$, then these two cuts together can have at most two unsafe edges ($G$ is $(p,2)$-flex-connected). This would contradict at least one of the equations \eqref{eq:a},\eqref{eq:b},\eqref{eq:c},\eqref{eq:d},\eqref{eq:e},\eqref{eq:f} since the number of unsafe edges in the LHS and RHS of these equations will not match. This is simple to see for equations \eqref{eq:c}-\eqref{eq:f}. For \eqref{eq:a} (or \eqref{eq:b}), we will obtain that the RHS has at most 4 unsafe edges (as $|E(A\setminus B, B\setminus A)| = 1$) in this case) while the LHS has at least 6 unsafe edges. With this observation, we can conclude that the four corner cuts have sizes $p+1,p+2,p+2,p+3$ in some permutation. In addition, the residual graph has no diagonal edges and so has only four residual edges as shown in figure \ref{fig:doublecrossing} below. Let $\alpha = (p+1)/2$. Equations \eqref{eq:c},\eqref{eq:d},\eqref{eq:e},\eqref{eq:f} further tell us that two of these residual edges must have cardinality $\alpha$ and must be adjacent and the other two must have cardinality $\alpha+1$. Furthermore, the residual edges of cardinality $\alpha$ define a $p+1$-cut and so must both be thin edges. These observations are summarized in Lemma \ref{lem:doublecrossing}.

    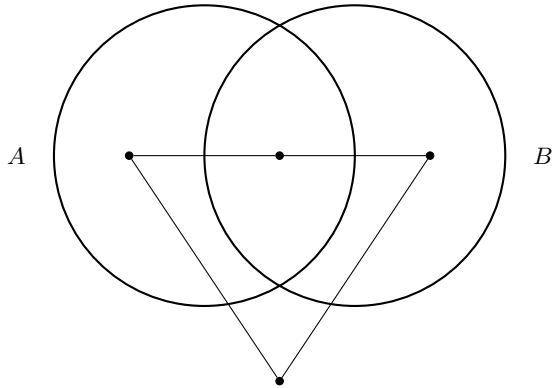
\begin{figure}[H]
    \centering
    \begin{tikzpicture}
        \draw[thick] (-2,0) circle (2cm);
        \draw[thick] (0,0) circle (2cm);
        \node[circle, draw, fill = black, inner sep=1pt] at (-1,0) (node y) {};
        \node[circle, draw, fill = black, inner sep=1pt] at (-3,0) (node x) {};
        \node[circle, draw, fill = black, inner sep=1pt] at (1,0) (node z) {};
        \node[circle, draw, fill = black, inner sep=1pt] at (-1,-3) (node w) {};
        \draw (node x) -- (node y);
        \draw (node y) -- (node z);
        \draw (node z) -- (node w);
        \draw (node x) -- (node w);
        \node at (-4.5,0) {$A$};
        \node at (2.5,0) {$B$};
    \end{tikzpicture}
    \caption{Residual graph of two crossing violated sets. Two of the edges have cardinality $\alpha$ and are adjacent and the other two have cardinality $\alpha +1$. The edges of size $\alpha$ are thin.}
    \label{fig:doublecrossing}
\end{figure}

\begin{lemma}\label{lem:doublecrossing}
    The residual graph of two crossing violated cuts does not contain diagonal edges and has four edges in total. A pair of adjacent edges have size $\alpha$ and are thin. The other pair of edges have size $\alpha+1$.
\end{lemma}

\begin{figure}[H]
    \centering
    \begin{tikzpicture}
        \draw[thick] (-2,0) circle (2cm);
        \draw[thick] (0,0) circle (2cm);
        \draw[thick] (-1,-2) circle (2cm);
        \node[circle, draw, fill = black, inner sep=1pt] at (-1,0.5) (node y) {}; \node at (-1,0.7) {\small (b)};
        \node[circle, draw, fill = black, inner sep=1pt] at (-3,0.5) (node x) {}; \node at (-3,0.7) {\small (a)};
        \node[circle, draw, fill = black, inner sep=1pt] at (1,0.5) (node z) {}; \node at (1,0.7) {\small (c)};
        \node[circle, draw, fill = black, inner sep=1pt] at (-1,-3.5) (node w) {}; \node at (-1,-3.7) {\small (g)};
        \node[circle, draw, fill = black, inner sep=1pt] at (-2,-1.5) (node v) {}; \node at (-2.2,-1.6) {\small (d)};
        \node[circle, draw, fill = black, inner sep=1pt] at (0,-1.5) (node u) {}; \node at (0.2,-1.6) {\small (f)};
        \node[circle, draw, fill = black, inner sep=1pt] at (-1,-1) (node t) {}; \node at (-1,-1.2) {\small (e)};
        \node[circle, draw, fill = black, inner sep=1pt] at (-4.5,-0.5) (node s) {}; \node at (-4.5,-0.3) {\small (h)};
        \node[circle, draw, fill = black, inner sep=1pt] at (1,-3.5) (node r) {}; \node at (1,-3.3) {\small (h)};
        \node[circle, draw, fill = black, inner sep=1pt] at (2.5,-0.5) (node q) {}; \node at (2.5,-0.3) {\small (h)};
        \draw (node x) -- (node y) -- (node z) -- (node u) -- (node w) -- (node v) -- (node t) -- (node y);
        \draw (node t) -- (node u); \draw (node v) -- (node x) -- (node s); \draw (node w) -- (node r); \draw (node z) -- (node q);
        \node at (-4.5,1.2) {$A$}; \node at (2.5,1.2) {$B$}; \node at (-1,-4.5) {$C$};
    \end{tikzpicture}
    \caption{Residual graph of three pair-wise crossing sets. Either residual vertices $(e)$ and $(h)$ are both empty or residual vertices $(g)$ and $(b)$ are both empty.}
    \label{fig:triplecrossing}
\end{figure}
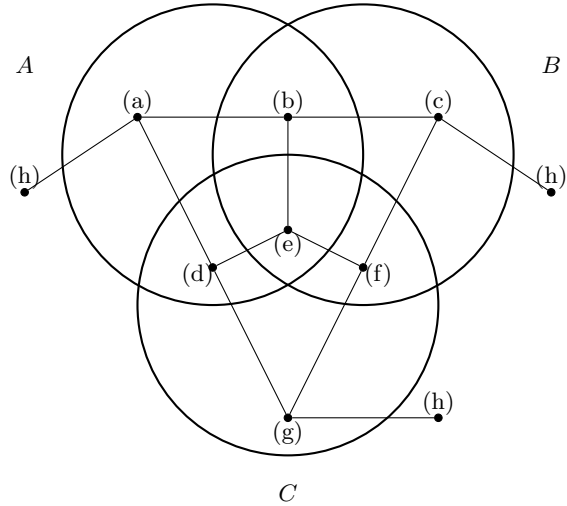

We next analyze the residual graph when three violated sets $A,B,C$ all cross pairwise. All possible edges are depicted in Figure \ref{fig:triplecrossing}. Any other edge is a diagonal edge for a particular pair of violated sets from $A,B,C$. We begin with a useful lemma. Note that for any node set $D$, either $D$ is empty or $\delta(D)$ has size at least $p$.

\begin{lemma}\label{lem:triplecrossinguseful}
    Let $A,B,C$ be any three violated sets. Suppose there exist four sets $D_1,D_2,D_3,D_4$ such that the following multiset equation holds: $\delta(A) + \delta(B) + \delta(C) = \delta(D_1) + \delta(D_2) + \delta(D_3) + \delta(D_4)$. Then, at least one of the four sets $D_1,D_2,D_3,D_4$ is empty.
\end{lemma}

\begin{proof}
    Suppose not, then the cuts $\delta(D_1),\delta(D_2),\delta(D_3),\delta(D_4)$ must have cardinality at least $p$ (since $G$ is $(p,2)$-flex-connected). Let these sizes be $p+w_1,p+w_2,p+w_3,p+w_4$. We immediately obtain $(p+2)+(p+2) + (p+2) = (p+w_1) + (p+w_2) + (p+w_3) + (p+w_4)$ and so $6 = p+w_1+w_2+w_3+w_4$. Therefore, the four cuts $\delta(D_1),\delta(D_2),\delta(D_3),\delta(D_4)$ do not have size larger than $6$. Since $p\geq 3$, $w_1+w_2+w_3+w_4 \leq 3$. Suppose $w_1<2$, the the corresponding cut $\delta(D_1)$ has at most $w_1$ unsafe edges (since $G$ is $(p,2)$-flex-connected). The same holds true for $w_2,w_3,w_4$. Now if all $w_1,w_2,w_3,w_4$ are strictly less than two, then there are at most three unsafe edges in total in the RHS of the equation in the lemma. But the LHS of the equation has at least 9 unsafe edges (since $A,B,C$ are violated sets). If any of $w_1,w_2,w_3,w_4$ is at least 2, say $w_1$ then the sum of the rest $w_2+w_3+w_4$ is at most 1 and so the corresponding cuts combined contain at most 7 unsafe edges (6 from the cut with size $p+w_1$ and 1 from the other three). This is again a contradiction.\qed
\end{proof}

We now show that at least one of the four residual vertices $(a),(c),(g),(e)$ must be empty. It is simple to verify the following using figure \ref{fig:triplecrossing}. Recall that there are no diagonal edges by Lemma \ref{lem:doublecrossing}.
\begin{align*}
    \delta(A) + \delta(B) + \delta(C) = \delta(a) + \delta(c) + \delta(g) + \delta(e)
\end{align*}
But then by Lemma \ref{lem:triplecrossinguseful}, we can conclude that either the residual vertex $(e)$ is empty and so $A\cap B\cap C$ is empty or the residual vertex $(g)$ is empty (by symmetry of the residual vertices $(a),(c),(g)$) and so $C\subseteq A\cup B$.

\textit{Case 1:} When $A\cap B\cap C$ is empty i.e. residual vertex $(e)$ is empty, we further argue that residual vertex $(h)$ is also empty. Notice that the residual vertices $(b),(d),(f)$ are all non-empty since the sets $A,B,C$ pair-wise cross. The following equation is easy to verify using figure \ref{fig:triplecrossing}:
\begin{align*}
    \delta(A) + \delta(B) + \delta(C) = \delta(b) + \delta(d) + \delta(f) + \delta(h)
\end{align*}
Then again by Lemma \ref{lem:triplecrossinguseful}, we can conclude that residual vertex $(h)$ must be empty.

\textit{Case 2:} When $C\subseteq A\cup B$ i.e. residual vertex $(g)$ is empty, we further argue that residual vertex $(b)$ is also empty. Notice that the residual vertices $(d),(f),(h)$ are all non-empty since the sets $A,B,C$ pair-wise cross. The following equation is easy to verify using figure \ref{fig:triplecrossing}:
\begin{align*}
    \delta(A) + \delta(B) + \delta(C) = \delta(b) + \delta(d) + \delta(f) + \delta(h)
\end{align*}
Then again by Lemma \ref{lem:triplecrossinguseful}, we can conclude that residual vertex $(b)$ must be empty. All these observations are summarized in Lemma \ref{lem:triplecrossing} and Figure \ref{fig:triplerefined}.

\begin{lemma}\label{lem:triplecrossing}
    Let $A,B,C$ be three violated sets that cross pair-wise. Then either (i) $A\cap B\cap C$ and $\overline{A\cup B\cup C}$ are empty or (ii) $A\cap B \subseteq C \subseteq A\cup B$ (after suitably permuting the labels $A,B,C$).
\end{lemma}
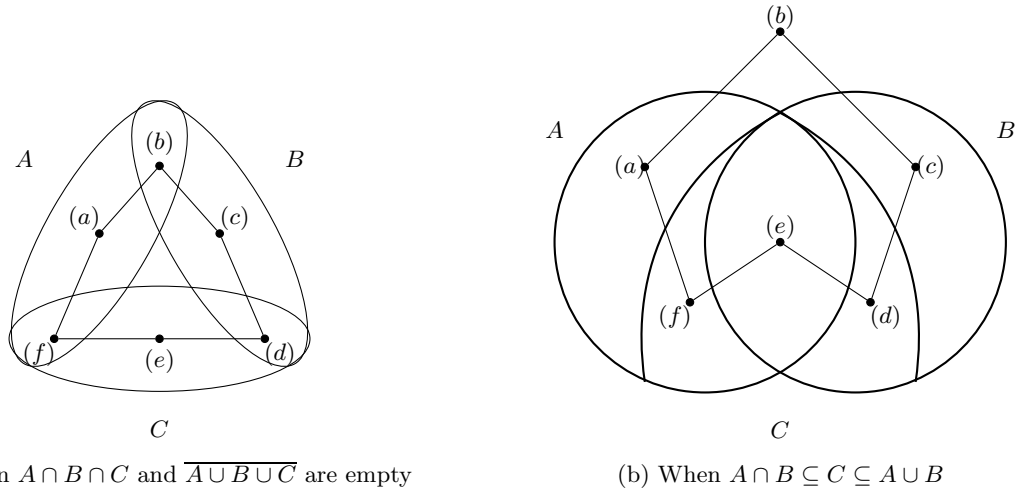
\begin{figure}[H]
    \centering
    \begin{subfigure}{0.5\textwidth}
        \centering
        \begin{tikzpicture}
            \draw (0,0) circle [x radius=2cm,y radius =0.7cm, rotate =60]; \node at (-1,1) {$A$};
            \draw (1.6,0) circle [x radius=2cm,y radius =0.7cm, rotate =120]; \node at (2.6,1) {$B$};
            \draw (0.8,-1.4) circle [x radius=2cm,y radius =0.7cm]; \node at (0.8,-2.6) {$C$};
            \node[circle, draw, fill = black, inner sep=1pt] at (0,0) (node z) {};\node at (-0.2,0.2) {\small $(a)$};
            \node[circle, draw, fill = black, inner sep=1pt] at (1.6,0) (node y) {};\node at (1.8,0.2) {\small $(c)$};
            \node[circle, draw, fill = black, inner sep=1pt] at (0.8,-1.4) (node x) {}; \node at (0.8,-1.7) {\small $(e)$};
            \node[circle, draw, fill = black, inner sep=1pt] at (-0.6,-1.4) (node w) {}; \node at (-0.8,-1.6) {\small $(f)$};
            \node[circle, draw, fill = black, inner sep=1pt] at (2.2,-1.4) (node v) {}; \node at (2.4,-1.6) {\small $(d)$};
            \node[circle, draw, fill = black, inner sep=1pt] at (0.8,0.9) (node u) {};\node at (0.8,1.2) {\small $(b)$};
            \draw (node u) -- (node z) -- (node w) -- (node x) -- (node v) -- (node y) -- (node u);
        \end{tikzpicture}
        \caption{When $A\cap B\cap C$ and $\overline{A\cup B\cup C}$ are empty}
        \label{fig:triplerefinedone}
    \end{subfigure}%
    \begin{subfigure}{0.5\textwidth}
        \centering
        \begin{tikzpicture}
            \draw[thick] (0,0) circle (2cm);\node at (-2,1.5) {$A$};
            \draw[thick] (2,0) circle (2cm);\node at (4,1.5) {$B$};
            \draw[bend right = 35,thick,] (1,1.72) to (-0.8,-1.86);
            \draw[bend left = 35,thick,] (1,1.72) to (2.8,-1.86); \node at (1,-2.5) {$C$};
            \node[circle, draw, fill = black, inner sep=1pt] at (1,2.8) (node a) {}; \node at (1,3) {\small $(b)$};
            \node[circle, draw, fill = black, inner sep=1pt] at (-0.8,1) (node b) {}; \node at (-1,1) {\small $(a)$};
            \node[circle, draw, fill = black, inner sep=1pt] at (-0.2,-0.8) (node c) {}; \node at (-0.4,-1) {\small $(f)$};
            \node[circle, draw, fill = black, inner sep=1pt] at (1,0) (node d) {}; \node at (1,0.2) {\small $(e)$};
            \node[circle, draw, fill = black, inner sep=1pt] at (2.2,-0.8) (node e) {}; \node at (2.4,-1) {\small $(d)$};
            \node[circle, draw, fill = black, inner sep=1pt] at (2.8,1) (node f) {}; \node at (3,1) {\small $(c)$};
            \draw (node a) -- (node b) -- (node c) -- (node d) -- (node e) -- (node f) -- (node a);
        \end{tikzpicture}
        \caption{When $A\cap B \subseteq C\subseteq A\cup B$}
        \label{fig:triplerefinedtwo}
    \end{subfigure}
    \caption{Residual graph of three pair-wise crossing violated sets after refinement}
    \label{fig:triplerefined}
\end{figure}

\begin{figure}[H]
    \centering
    \begin{tikzpicture}
        \draw[thick] (0,0) circle (2cm); \node at (-2,1.5) {$A$};
        \draw[thick] (2,0) circle (2cm); \node at (4,1.5) {$B$};
        \draw[thick] (2,0) circle (1cm); \node at (3,-1) {$C$};
        \node[circle, draw, fill = black, inner sep=1pt] at (-1,0) (node a) {}; \node at (-1,-0.3) {\small $(a)$};
        \node[circle, draw, fill = black, inner sep=1pt] at (0.5,0) (node b) {}; \node at (0.5,-0.3) {\small $(b)$};
        \node[circle, draw, fill = black, inner sep=1pt] at (1.5,0) (node c) {}; \node at (1.5,-0.3) {\small $(c)$};
        \node[circle, draw, fill = black, inner sep=1pt] at (2.5,0) (node d) {}; \node at (2.5,-0.3) {\small $(d)$};
        \node[circle, draw, fill = black, inner sep=1pt] at (3.5,0) (node e) {}; \node at (3.5,-0.3) {\small $(e)$};
        \node[circle, draw, fill = black, inner sep=1pt] at (1,3) (node f) {}; \node at (1,3.3) {\small $(f)$};
        \draw (node a) -- (node b) -- (node c) -- (node d) -- (node e) -- (node f) -- (node a);
        \draw[bend right = 60,dashed] (node a) to (node c);
        \draw[bend right = 60,dashed] (node b) to (node e);
        \draw (node d) -- (node f);        
    \end{tikzpicture}
    \caption{Residual graph of three violates sets where $A$ crosses $B$ and $C$, and $C\subseteq B$. Either $A$ crosses both $B$ and $C$ amenably or crosses none of them amenably.}
    \label{fig:triplecontainedrossing}
\end{figure}
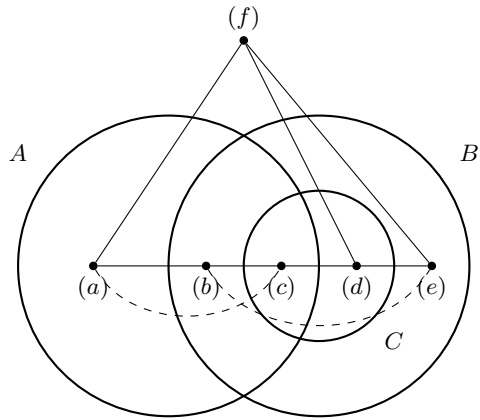

Lastly, we analyze the structure of three violated cuts $A,B,C$ where $C\subseteq B$ and $A$ crosses both $B$ and $C$. All possible edges are depicted in Figure \ref{fig:triplecontainedrossing}. Any other edge is a diagonal edge for a particular pair of crossing sets among $A,B,C$.

Let us look at the set $A$ and its crossings with $B$ and $C$ individually. In the crossing with $B$, the residual edges $(be)$ and $(cd)$ are the internal edges of the cut $\delta(A)$ and the residual edge $(af)$ is the external edge. We know from Lemma \ref{lem:doublecrossing} that one of these edges has size $\alpha$ and is thin and the other has size $\alpha + 1$. Thus, $|(af)| \geq \alpha$ and $|(be)| + |(cd)| \leq \alpha+1$. In the crossing with $C$, the residual edges $(be)$ and $(af)$ are the external edges of the cut $\delta(A)$ and the residual edge $(cd)$ is the internal edge. Thus $|(cd)|\geq \alpha$. This implies that $|(be)|\leq 1$. Now if $|(be)| = 1$, then both the residual edges $|(af)|$ and $|(cd)|$ must have size $\alpha$ and must be thin. Since $\delta(A)$ has at least three unsafe edges each of the residual edges $(be), (af)$ and $(cd)$ must contain exactly one unsafe edge each. In this case, $A$ does not cross either of $B$ or $C$ amenably. On the other hand, if $|(be)| = 0$, then the external (internal) edges of $\delta(A)$ with respect to both the sets $B$ and $C$ are the same. Hence, either $A$ crosses both $B$ and $C$ amenably or crosses neither of them amenably. These observation are summarised in Lemma \ref{lem:triplecontaincrossing} below.

\begin{lemma}\label{lem:triplecontaincrossing}
    Let $A,B,C$ be three violates sets such that $A$ crosses both $B$ and $C$ and $C\subseteq B$. Then either $A$ crosses both $B$ and $C$ amenably or $A$ crosses neither of them amenably.
\end{lemma}

With these structures in place, we can now prove the following lemma about amenable crossings.

\begin{lemma}\label{lem:amenablecrossing}
    Suppose a violated set $A$ crosses another violated set $B$ amenably. Then for any violated set $C$, either $A$ does not cross $C$ or $A$ crosses $C$ amenably.
\end{lemma}

\begin{proof}
    Suppose not, then there exists a triple of violated sets $A,B,C$ such that $A$ crosses $B$ amenably and $A$ crosses $C$ but not amenably. We consider three cases (each with sub-cases).
    \begin{description}
        \item[Case 1:] $B$ crosses $C$. Then, one of the cases shown in figure \ref{fig:triplerefined} occurs. In the first case shown in Figure \ref{fig:triplerefinedone}, the internal edge $(bc)$ of $\delta(A)$ with respect to the crossing with $B$ is the external edge of $\delta(A)$ with respect to the crossing with $C$, and the external edge $(ef)$  of $\delta(A)$ with respect to the crossing with $B$ is the internal edge of $\delta(A)$ with respect to the crossing with $C$. Hence $A$ must necessarily cross $C$ amenably. In the second case, shown in Figure \ref{fig:triplerefinedtwo}, the internal edge $(de)$ of $\delta(A)$ with respect to the crossing with $B$ is the internal edge of $\delta(A)$ with respect to the crossing with $C$, and the external edge $(ab)$  of $\delta(A)$ with respect to the crossing with $B$ is the external edge of $\delta(A)$ with respect to the crossing with $C$. Hence $A$ must necessarily cross $C$ amenably. \\
        \item[Case 2:] $B$ does not cross $C$ but $B\cup C = V$. This implies that $A \subseteq B\cup C$. We refer the reader to Figure \ref{fig:triplecrossing} for this case where now residual vertices $(a)$ and $(h)$ are empty. The internal edges of $\delta(A)$ with respect to $B$ are $(bc)$ and $(ef)$ and the external edge is $(dg)$. Thus $|(dg)|\geq \alpha$ and $|(bc)|+|(ef)| \leq \alpha+1$. The internal edges of $\delta(A)$ with respect to $C$ are $(de)$ and $(ef)$ and the external edge is $(bc)$. Thus $|(bc)| \geq \alpha$ and so $|(ef)|\leq 1$. Now if $|(ef)| = 1$, then $|(bc)| = |(de)| = \alpha$ and these are thin edges. Since $\delta(A)$ has at least three unsafe edges, each of the residual edges $(bc),(de),(ef)$ contain exactly one unsafe edge each. This contradicts the assumption that $A$ crosses $B$ amenably. On the other hand, if $|(ef)| = 0$, then the internal (external) edge of $\delta(A)$ with respect to $B$ is the external (internal) edge of $\delta(A)$ with respect to $C$ and so $A$ crosses $C$ amenably. \\
        \item[Case 3:] $B$ does not cross $C$ and $B\cup C \neq V$. Then either $B$ and $C$ are disjoint or one is contained in the other. If one is contained in the other, then Lemma \ref{lem:triplecontaincrossing} shows that $A$ crosses $C$ amenably. If $B$ and $C$ are disjoint, then the residual graph is given in figure \ref{fig:tripledisjointcrossing} with only the edges corresponding to $\delta(A)$.
        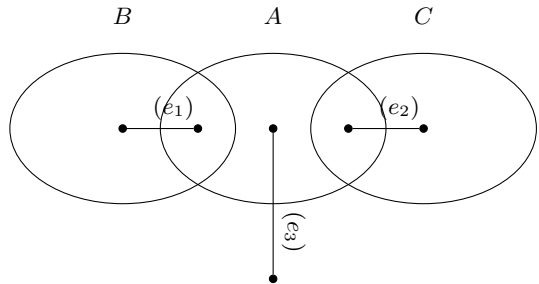
\begin{figure}[H]
        \centering
        \begin{tikzpicture}
            \draw (0,0) ellipse (1.5cm and 1cm);
            \draw (2,0) ellipse (1.5cm and 1cm);
            \draw (4,0) ellipse (1.5cm and 1cm);
            \node at (0,1.5) {$B$};
            \node at (2,1.5) {$A$};
            \node at (4,1.5) {$C$};
            \node[circle, draw, fill = black, inner sep=1pt] at (0,0) (node z) {};
            \node[circle, draw, fill = black, inner sep=1pt] at (2,0) (node y) {};
            \node[circle, draw, fill = black, inner sep=1pt] at (4,0) (node x) {};
            \node[circle, draw, fill = black, inner sep=1pt] at (2,-2) (node w) {};
            \node[circle, draw, fill = black, inner sep=1pt] at (1,0) (node v) {};
            \node[circle, draw, fill = black, inner sep=1pt] at (3,0) (node u) {};
            \draw (node z) -- (node v) node[above,pos=0.7,sloped] {$(e_1)$};
            \draw (node u) -- (node x) node[above,pos=0.7,sloped] {$(e_2)$};
            \draw (node y) -- (node w) node[above,pos=0.7,sloped] {$(e_3)$};
        \end{tikzpicture}
        \caption{$A$ crosses $B$ and $C$ and $B\cap C = \emptyset$. The residual edges $(e_1),(e_2),(e_3)$ of $\delta(A)$ are shown.}
        \label{fig:tripledisjointcrossing}
    \end{figure}
    The external edges of $\delta(A)$ with respect to $B$ are $(e_2)$ and $(e_3)$ and the internal edge is $(e_1)$. Thus $|(e_1)|\geq \alpha$ and $|(e_2)|+|(e_3)| \leq \alpha+1$. The external edges of $\delta(A)$ with respect to $C$ are $(e_1)$ and $(e_3)$ and the internal edge is $(e_2)$. Thus $|(e_2)| \geq \alpha$ and so $|(e_3)|\leq 1$. Now if $|(e_3)| = 1$, then $|(e_1)| = |(e_2)| = \alpha$ and these are thin edges. Since $\delta(A)$ has at least three unsafe edges, each of the residual edges $(e_1),(e_2),(e_3)$ contain exactly one unsafe edge each. This contradicts the assumption that $A$ crosses $B$ amenably. On the other hand, if $|(e_3)| = 0$, then the internal (external) edge of $\delta(A)$ with respect to $B$ is the external (internal) edge of $\delta(A)$ with respect to $C$ and so $A$ crosses $C$ amenably.
    \end{description}\qed
\end{proof}

\begin{lemma}\label{lem:amenableuncrossable}
    Let $\mathcal{F}_1$ be the family of violated sets that either (i) do not cross any other violated set or (ii) cross some violated set amenably. Then $\mathcal{F}_1$ forms an uncrossable family.
\end{lemma}

\begin{proof}
    Let $A$, $B$ be two crossing violated sets in the family $\mathcal{F}_1$. Due to the lemma \ref{lem:amenablecrossing}, both the sets $A$ and $B$ cross each other amenably. We consider three cases:
    \begin{description}
        \item[Case 1:] The external edges of $\delta(A)$ and $\delta(B)$ do not contain any unsafe edges. Due to lemma \ref{lem:doublecrossing}, the internal edges of both the cuts, contain at least three unsafe edges each and must have cardinality $\alpha+1$. The external edges must have cardinality $\alpha$. Hence, the sets $A\setminus B$ and $B\setminus A$ are also violated (the corresponding cuts are $(p+2)$-cuts with at least three unsafe edges). We will show that $A\setminus B$ (and by symmetry $B\setminus A$) is in $\mathcal{F}_1$. Suppose not, then there exists a set $C \in \mathcal{F}_1$ that $A\setminus B$ crosses but not amenably. We know that all unsafe edges in $\delta(A\setminus B)$ have one end point in $A\setminus B$ and the other end point in $A\cap B$. At least one of these unsafe edges is an external edge in the crossing between $A\setminus B$ and $C$ and at least one is an internal edge. Thus there exists an unsafe edge $(a_1,b_1)$ with $a_1 \in (A\setminus B) \cap C$ and $b_1 \in A\cap B \cap C$ and an unsafe edge $(a_2,b_2)$ with $a_2 \in (A\setminus (B\cup C))$ and $b_2\in (A\cap B)\setminus C$. This implies that $B$ crosses $C$ since $a_1 \in C\setminus B$, $b_1 \in B\cap C$, $a_2\in \overline{B\cup C}$ and $b_2\in B\setminus C$. But now, the edge $(a_1,b_1)$ is an unsafe internal edge in the crossing between $B$ and $C$ and the edge $(a_2,b_2)$ is an unsafe external edge in the crossing between $B$ and $C$. This implies that $B$ crosses $A$ amenably but does not cross $C$ amenably contradicting lemma \ref{lem:amenablecrossing}.\\
        \item[Case 2:] The internal edges of $\delta(A)$ and $\delta(B)$ do not contain any unsafe edges. Due to lemma \ref{lem:doublecrossing}, the external edges of both the cuts, contain at least three unsafe edges each and must have cardinality $\alpha+1$. The internal edges must have cardinality $\alpha$. Hence, the sets $A\setminus B$ and $B\setminus A$ are also violated (the corresponding cuts are $(p+2)$-cuts with at least three unsafe edges). We will show that $A\setminus B$ (and by symmetry $B\setminus A$) is in $\mathcal{F}_1$. Suppose not, then there exists a set $C \in \mathcal{F}_1$ that $A\setminus B$ crosses but not amenably. We know that all unsafe edges in $\delta(A\setminus B)$ have one end point in $A\setminus B$ and the other end point in $\overline{A\cup B}$. At least one of these unsafe edges is an external edge in the crossing between $A\setminus B$ and $C$ and at least one is an internal edge. Thus there exists an unsafe edge $(a_1,b_1)$ with $a_1 \in (A\setminus B) \cap C$ and $b_1 \in \overline{A\cup B} \cap C$ and an unsafe edge $(a_2,b_2)$ with $a_2 \in (A\setminus (B\cup C))$ and $b_2\in \overline{A\cup B\cup C}$. This implies that $A$ crosses $C$ since $a_1 \in A\cap C$, $b_1 \in C\setminus A$, $a_2\in A\setminus C$ and $b_2\in \overline{A\cup C}$. But now, the edge $(a_1,b_1)$ is an unsafe internal edge in the crossing between $A$ and $C$ and the edge $(a_2,b_2)$ is an unsafe external edge in the crossing between $A$ and $C$. This implies that $A$ crosses $B$ amenably but does not cross $C$ amenably contradicting lemma \ref{lem:amenablecrossing}.\\
        \item[Case 3:] The internal edge of $\delta(A)$ does not contain any unsafe edges and the external edge of $\delta(B)$ does not contain any safe edges. Due to lemma \ref{lem:doublecrossing}, the external (internal) residual edge of $\delta(A)$ ($\delta(B)$) contains at least three unsafe edges and must have cardinality $\alpha+1$. The other residual edges must have cardinality $\alpha$. Hence, the sets $A\cup B$ and $A\cap B$ are also violated (the corresponding cuts are $(p+2)$-cuts with at least three unsafe edges). 
        
        Suppose there exists a set $C$ such that $A\cup B$ crosses $C$ but not amenably. We know that all unsafe edges in $\delta(A\cup B)$ have one end point in $A\setminus B$ and the other end point in $\overline{A\cup B}$. At least one of these unsafe edges is an external edge in the crossing between $A\cup B$ and $C$ and at least one is an internal edge. Thus there exists an unsafe edge $(a_1,b_1)$ with $a_1 \in (A\setminus B) \cap C$ and $b_1 \in \overline{A\cup B} \cap C$ and an unsafe edge $(a_2,b_2)$ with $a_2 \in (A\setminus (B\cup C))$ and $b_2\in \overline{A\cup B\cup C}$. This implies that $A$ crosses $C$ since $a_1 \in A\cap C$, $b_1 \in C\setminus A$, $a_2\in A\setminus C$ and $b_2\in \overline{A\cup C}$. But now, the edge $(a_1,b_1)$ is an unsafe internal edge in the crossing between $A$ and $C$ and the edge $(a_2,b_2)$ is an unsafe external edge in the crossing between $A$ and $C$. This implies that $A$ crosses $B$ amenably but does not cross $C$ amenably contradicting lemma \ref{lem:amenablecrossing}. Thus $A\cup B \in \mathcal{F}_1$. 
        
        Suppose there exists a set $C$ such that $A\cap B$ crosses $C$ but not amenably. We know that all unsafe edges in $\delta(A\cap B)$ have one end point in $A\setminus B$ and the other end point in $A\cap B$. At least one of these unsafe edges is an external edge in the crossing between $A\cap B$ and $C$ and at least one is an internal edge. Thus there exists an unsafe edge $(a_1,b_1)$ with $a_1 \in (A\setminus B) \cap C$ and $b_1 \in A\cap B \cap C$ and an unsafe edge $(a_2,b_2)$ with $a_2 \in (A\setminus (B\cup C))$ and $b_2\in (A\cap B)\setminus C$. This implies that $B$ crosses $C$ since $a_1 \in C\setminus B$, $b_1 \in B\cap C$, $a_2\in \overline{B\cup C}$ and $b_2\in B\setminus C$. But now, the edge $(a_1,b_1)$ is an unsafe internal edge in the crossing between $B$ and $C$ and the edge $(a_2,b_2)$ is an unsafe external edge in the crossing between $B$ and $C$. This implies that $B$ crosses $A$ amenably but does not cross $C$ amenably contradicting lemma \ref{lem:amenablecrossing}.\\ 
    \end{description}\qed
\end{proof}
\begin{lemma}\label{lem:nonamenableuncrossable}
    Let $\mathcal{F}$ be the family of violated cuts and $\mathcal{F}_1$ be the family of violated cuts from lemma \ref{lem:amenableuncrossable}. Then the family of violated cuts $\mathcal{F}_2 = \mathcal{F}\setminus \mathcal{F}_1$ forms an uncrossable family.
\end{lemma}

\begin{proof}
    Let $A$, $B$ be two violated sets in $\mathcal{F}_2$. Recall that $A$ and $B$ are $(p+2)$-cuts with at least 3 unsafe edges each. Suppose $A$ and $B$ cross, then they do not cross each other amenably. Referring back to Lemma \ref{lem:doublecrossing}, both the thin edges in their crossing must have an unsafe edge each, but this means that there exists a $p+1$-cut with at least 2 unsafe edges contradicting the assumption that $G$ is $(p,2)$-flex-connected. Hence no pair of sets $A$ and $B$ in $\mathcal{F}_2$ cross each other.\qed
\end{proof}

The upshot of Lemmas \ref{lem:amenableuncrossable} and \ref{lem:nonamenableuncrossable} is that we can start with a $(7+\epsilon)$-approximate $(p,2)$-flex connected solution using the results in \cite{nutov2023improved} and then augment the violated cuts in $\mathcal{F}_1$ and $\mathcal{F}_2$ using the WGMV algorithm. The size of the family $\mathcal{F}$ is polynomial as shown in \cite{boyd2024approximation} and so the families $\mathcal{F}_1$ and $\mathcal{F}_2$ can be found in polynomial time. Since these families are uncrossable, an additional factor of 4 is incurred providing an $(11+\epsilon)$-approximation algorithm for $(p,3)$-FGC when $p$ is odd. This completes the proof of Theorem \ref{thm:FGC}. \qed

\vfill
\clearpage

\bibliographystyle{splncs04}
\bibliography{arxiv}

\vfill
\clearpage

\appendix

\centerline{\Huge Appendix}

\section{Tightness of Theorem \ref{thm:crossdensity}}\label{apx:tightresult}

In this section, we argue that the result in Theorem \ref{thm:crossdensity} is nearly asymptotically tight. We do so by exhibiting a family of instances where the crossing density is $O(|V|)$ while the WGMV primal-dual algorithm provides an approximation ratio of $\Omega(\sqrt{|V|})$. This example was found by BCGI \cite{bansal2024improved} and we borrow the ideas from them.

\begin{figure}[H]
\centering
\begin{subfigure}{.5\textwidth}
  \centering
  \includegraphics[width=1\linewidth]{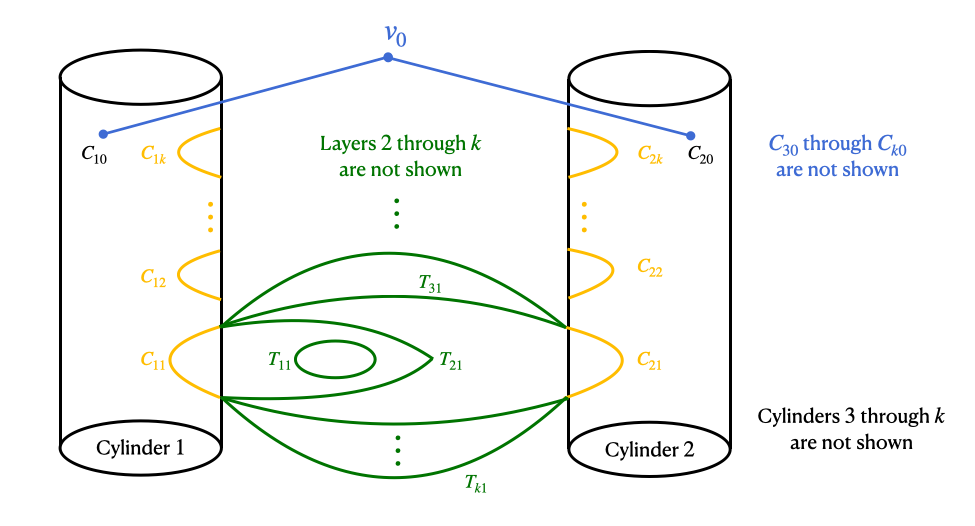}
  \caption{Full View}
\end{subfigure}%
\begin{subfigure}{.5\textwidth}
  \centering
  \includegraphics[width=1\linewidth]{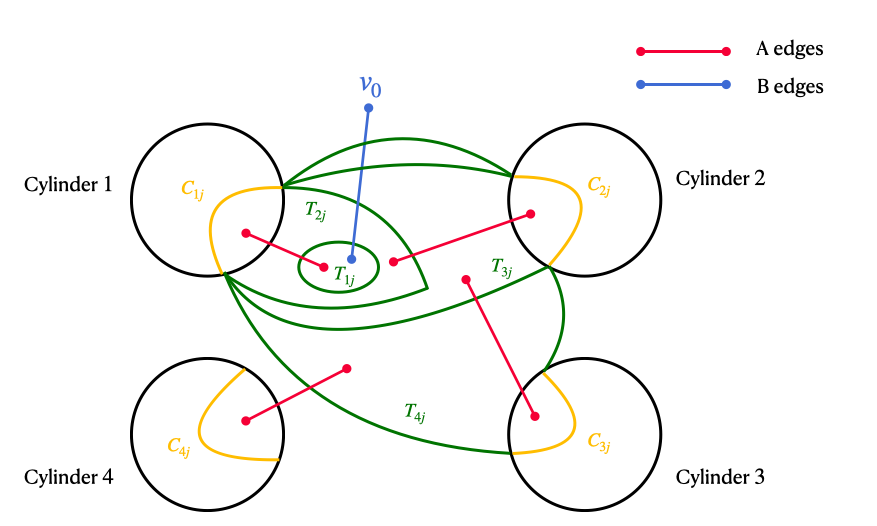}
  \caption{Cross Section with edges}
\end{subfigure}
\caption{Tight example for Theorem \ref{thm:crossdensity}}
\label{fig:WGMV_counterexample}
\end{figure}

We have node-sets $C_{ij}$ for
$i=1,2,\ldots,k$ and $j=0,1,2,\ldots,k$. All these $C$-sets are
pair-wise disjoint. Additionally, for $j'=1,\ldots, k$, we have
node-sets $T_{1j'} \subsetneq T_{2j'}\subsetneq \cdots T_{kj'}$ and
each of these are disjoint from all the $C$-sets.  Additionally,
we have at least one node $v_0$ lying outside the union of all these
$C$-sets and $T$-sets. See Figure~\ref{fig:WGMV_counterexample}.
We designate $\mathcal{F'}$ to be the following family of node-sets
that consists of two types of sets:
\begin{align*}
& (I)  \qquad & C_i := \bigcup_{j=0}^k C_{ij} \quad               & \text{ for } i\in[k] \\
& (II) \qquad & T_{i'j'} \cup \bigcup_{(i,j) \in R'} C_{ij} \quad & \text{ for }
		i' \in[k],~ j'\in[k] \text{ and } \\
&&&		R' \subseteq R(i',j') \text{ where }
		R(i',j') = \{(i,j) \:|\: 1\leq i < i',\; 0\leq j\leq k\}
\end{align*}
Informally speaking, the sets $C_1,C_2,\dots,C_k$ can be viewed as
(pairwise-disjoint) ``cylinders'', see Figure~\ref{fig:WGMV_counterexample},
the (first) index $i$ is associated with one of these cylinders, and
note that $C_{i0} = C_i \setminus (\cup_{j=1}^k C_{ij})$;
the (second) index $j$ is associated with a ``layer'' (i.e., a horizontal plane), and
the sub-family
	$T_{1j'} \subsetneq T_{2j'}\subsetneq \cdots T_{kj'}$
	forms a nested family on layer~$j'$, see Figure~\ref{fig:WGMV_counterexample}.
For notational convenience, let $T_{0j'}=\emptyset,\; \forall j'\in[k]$.
Observe that a set of type~(II) is the union of one set $T_{i'j'}$ of
the nested family of layer~$j'$
together with the sets of an arbitrary sub-family of each of the
``cylinder families'' $\{ C_{i0}, C_{i1}, C_{i2}, \dots, C_{ik} \}$
with (first) index $i \leq i'$. The edges of the graph are of two types and have unit costs:

\begin{enumerate}
\item[(A)] For $i' = 1,\ldots,k$ and $j' = 1,\ldots, k$, we have
an edge from $T_{i'j'} \setminus T_{(i'-1)j'}$ to $C_{ij}$ where $i=i'$ and $j=j'$.

\item[(B)] For $j=1,\ldots, k$, we have an edge from $v_0$ to $T_{1j}$.
For $i=1,\ldots,k$, we have an edge from $v_0$ to $C_{i0}$.
\end{enumerate}

BCGI had shown that the WGMV primal-dual algorithm outputs a solution whose cost is $\Omega(k) = \Omega(\sqrt{|V|)}$ away from the optimal cost. It is easy to verify that the crossing density of the above example is $O(k^2) = O(|V|)$. Hence Theorem \ref{thm:crossdensity} can only be improved to an $O(\sqrt{\rho})$-approximation factor and no further.

\section{Extensions beyond Pliable Families}\label{apx:beyondpliable}

In this section we consider a further natural generalization of the class of pliable families. Pliability requires that at least two of the four corner cuts $A\cup B,A\cap B, A\setminus B, B\setminus A$ also lies in the family $\mathcal{F}$ whenever $A$ and $B$ lie in the family $\mathcal{F}$. We can relax this condition further to require at least one of the four corner cuts (instead of two). We show that the $\mathcal{F}$-augmentation problem of such families can capture the hitting set problem and so there is an $\Omega(\log |V|)$-hardness of approximation. A similar reduction was shown by BCGI \cite{bansal2024improved} to prove that the $\mathcal{F}$-augmentation problem in its most general form captures the hitting set problem.

Consider an instance of the hitting set problem with universe of elements $I=\{1,2,\ldots,n\}$ and sets to hit $T_1,\ldots,T_m \subseteq I$. Construct a graph $G$ with node set $V= \{a_1,\ldots,a_n,b_1,\ldots,b_n\}$ and edge set $E = \{(a_1,b_1),\ldots,(a_n,b_n)\}$ with unit costs i.e. $G$ is simply a matching between the $a$ and $b$ nodes. For each $T_j$, define $S_j = \{a_i : i\in T_j\}$. Define the family of cuts $\mathcal{F} := \{S_1,\ldots,S_m\}$. The hitting set problem is exactly captured by the $\mathcal{F}$-augmentation problem for the instance we created. However, $\mathcal{F}$ may not satisfy the property that at least one of the four corner cuts is also in the family. 

Observe that the hitting set problem remains unchanged upon taking unions of the sets $T_i$ since if $T_i$ and $T_j$ are `hit', then so is $T_i \cup T_j$. Thus, while defining $\mathcal{F}$, we can throw in all unions of the sets $S_j$ as well and the problem remains unchanged. This completes the reduction. Note that one could instead take the complements of the sets $S_j$ and take intersections instead of union to obtain the same reduction (due to De Morgan's Laws).

\section{Families with exactly one of bounded crossing density and property $\gamma$}\label{appendix:crossversusgamma}

In this section we exhibit two pliable set families. One of them has a bounded crossing density but does not satisfy property $\gamma$ while the other satisfies property $\gamma$ but does not have bounded crossing density.

\subsubsection{Bounded Crossing Density}~

    Consider a five vertex graph $v_1,v_2,v_3,v_4,v_5$. Define $C = \{v_1,v_2\}$, $S_1 = \{v_2,v_3\}$ and $S_2 = \{v_2,v_3,v_4\}$. Let the family $\mathcal{F}$ comprise of $C,S_1,S_2, C\cup S_1, S_1\setminus C, C\cup S_2$ and $S_2\setminus C$. Then we see that $C$ is an inclusion-wise minimal set in $\mathcal{F}$ which crosses $S_1\subseteq S_2$. The crossing density is clearly bounded since the number of sets in the family itself is bounded by 7. However, the set $S_2\setminus (S_1\cup C)$ is neither empty nor a member of $\mathcal{F}$ and so property $\gamma$ does not hold.

\subsubsection{Property $\gamma$}~

    This example is a bit more abstract. Consider a chain of sets $S_1\subset S_2 \subset \cdots \subset S_k$. Consider disjoint active sets $C_1,C_2,\ldots, C_k$ each of which cross the entire chain $S_1,\ldots, S_k$. Suppose there also exists an active set in each of the sets $S_i\setminus (S_{i-1} \cup C_1 \cup \cdots\cup C_k)$. Then, property $\gamma$ holds but the crossing density is $\Omega(k)$.

\section{Proof of Lemma \ref{lem:colornodes}}\label{appendix:lemproof}

We provide the proof of Lemma \ref{lem:colornodes} as given in \cite{bansal2024improved}.

\begin{proof}
\begin{enumerate}
    \item[(i)] Consider two sets $C_1,C_2 \in \mathcal{C}$ such that $C_1\cap{C_2}\neq\emptyset$. Then by the definition of pliable families,
one of the sets
$C_1\cap C_2$, $C_1\setminus{C_2}$, or $C_2\setminus{C_1}$ is violated. Thus,
a proper subset of either $C_1$ or $C_2$ is violated.  This is a
contradiction because $C_1$ and $C_2$ are minimal violated sets.
\item[(ii)] A leaf node of $\mathcal{T} $corresponds to a witness set $S$ that does not contain any other witness sets. However, it must contain an active set (that could be $S$) and now $S$ is the smallest witness set containing this active set.
\item[(iii)] We have $n_G \leq n_L$ because the number of leaves in
any tree is at least the number of nodes that are incident to three
or more edges of the tree.
Moreover, by~(ii), we have $n_L \leq n_R$. 
Every red node of $\mathcal{T}$ is associated with an active set,
hence, $n_R \leq |\mathcal{C}|$.
\end{enumerate}
\end{proof}

\end{document}